# Evaluating aliphatic CF, CF$_2$ and CF$_3$ groups as vibrational Stark effect reporters

R. Cruz,[1] K. Ataka,[1] J. Heberle,[1,2,*] and J. Kozuch [1,2,*]

[1] *Fachbereich Physik, Freie Universität Berlin, Berlin, 14195,* Germany

[2] *Forschungsbau SupraFAB, Freie Universität Berlin, Berlin, 14195,* Germany

Given the extensive use of fluorination in molecular design, it is imperative to understand the solvation properties of fluorinated compounds and the impact of the C-F bond on electrostatic interactions. Vibrational spectroscopy can provide direct insight into these interactions by using the C-F bond stretching (v(C-F)) as an electric field probe through the vibrational Stark effect (VSE). In this work, we explore the VSE of the three basic patterns of aliphatic fluorination, i.e., mono-, di-, and trifluorination in CF, CF$_2$ and CF$_3$ groups, respectively, and compare their response to the well-studied aromatic v(C-F). Magnitudes (i.e. Stark tuning rates) and orientations of the difference dipole vectors of the v(C-F)-containing normal modes were determined using density functional theory and a molecular dynamics (MD)-assisted solvatochromic analysis of model compounds in solvents of varying polarity. We obtain Stark tuning rates of 0.2 – 0.8 cm$^{-1}$/(MV/cm), with smallest and largest electric field sensitivities for CF$_{aliphatic}$ and CF$_{3,aliphatic}$, respectively. While average electric fields of solvation were oriented along the main symmetry axis of the CF$_n$, and thus along its static dipole, the Stark tuning rate vectors were tilted by up to 87° potentially enabling to map electrostatics in multiple dimensions. We discuss the influence of conformational heterogeneity on spectral shifts and point out the importance of multipolar and/or polarizable MD force fields to describe the electrostatics of fluorinated molecules. The implications of this work are of direct relevance for studies of fluorinated molecules as found in pharmaceuticals, fluorinated peptides and proteins.

**I. INTRODUCTION**

Fluorination is a very popular strategy to fine-tune or even drastically modify molecular properties via a structurally subtle chemical modification of C-H for C-F bonds.[1] As such, it has found many applications in agrochemistry, material sciences, and medicinal chemistry with examples of fluorinated drugs such as the 5α-reductase inhibitor dutasteride or the anti-HIV agent maraviroc.[2–4] Moreover, the last decade has seen increased interest in the use of fluorinated amino acids for protein engineering to (re)design the structure and function of proteins and enzymes, and by this contribute to further expand the amino acid tool box.[5],[6],[7] The extensive use of fluorinated substances has also led to a growing concern about their toxicity and environmental impact. Specifically, per- and poly-fluoroalkyl substances (PFAS) are persistent soil contaminants known to have adverse effects on human health[8–12].

The C-F bond has several distinctive properties, the most straightforward being that it is one of the strongest chemical bonds ensuring inertness in biologically and chemically aggressive environments. In addition, due to its small size, the F atom can be considered an isosteric substitution when replacing H atoms in molecular design approaches. However, the replacement impacts the electrostatic properties of the substituted molecule[1],[13] since C-F bonds are highly polar but hardly polarizable. As expected from the inclusion of a polar bond, monofluorination and trifluoromethylation of aliphatic molecules increases their water affinity.[13] Instead, additional fluorination leads to the opposite effect, resulting in the hydrophobicity, lipophilicity, and self-affinity of perfluorinated molecules.[14] Such hydrophobicity tuning is exploited in drug design to increase the affinity of molecules towards a specific pharmacological target.[15] The mechanism behind the paradoxical hydrophobicity observed with increasing fluorination is still under debate but was reported to involve a local electrostatic attraction, caused by the electric dipole of the C-F bond and steric hindrance within the hydrogen-bonded network of the hydration shell.[16],[17] With more than seven CF$_2$ groups, a van der Waals-type force has been proposed as the driving mechanism behind the so-called perfluorophilic interaction.[14,18]

Unveiling the role of electrostatics is fundamental to the understanding of the interactions occurring among fluorinated molecules and with their solvation environment. An experimental approach that can provide direct insight into the electrostatics underlying such non-covalent interactions is vibrational Stark effect (VSE) spectroscopy.[19] According to the phenomenon of VSE, the frequency of molecular vibrations ($\bar{v}$; in units of cm$^{-1}$) will shift due to the interaction of an electric field, $\vec{F}$ (units of MV/cm), with the difference dipole, $\Delta\vec{\mu}$ (units of cm$^{-1}$/(MV/cm)). The latter is the change in dipole moment between vibrational ground and excited states. Its magnitude $|\Delta\vec{\mu}|$ is often referred to as the Stark tuning rate, as it describes the sensitivity of a vibrational probe to electric fields. It is worth noting that this vector differs from the molecular dipole moment, $\vec{M}$, and the transition dipole moment, $\overrightarrow{TDM}$, which account for the orientation of solvent electric fields and the peak intensity in spectra, respectively (FIG. 1). In few cases, also the quadratic VSE becomes relevant, which depends on the difference polarizability tensor, $\Delta\underline{\alpha}$ (units of cm$^{-1}$/ (MV/cm)$^2$). Overall, this effect is expressed via the VSE equation

$$\bar{\nu} = \bar{\nu}_0 - \Delta\vec{\mu} \cdot \vec{F} - \frac{1}{2}\vec{F} \cdot \Delta\underline{\alpha} \cdot \vec{F} \qquad (1)$$

relative to the zero-field vibrational frequency, $\bar{\nu}_0$, which refers to the vibrational probe in vacuum. The most prominent VSE probes are C=O and C≡N oscillators,[19,20] which due to their small size have been used to quantify non-covalent, local, electrostatic interactions in many settings such as in solvents,[21–26] at electrode interfaces,[27–29] and in biological systems.[30–39] These studies enhanced our understanding of the role of electrostatics in many chemical, biochemical and biophysical processes; most notably the direct demonstration that electric fields can exert catalytic effects to drive (bio)chemical reactions.[40,41]

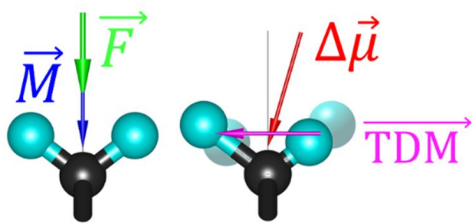

FIG. 1. Schematic depiction of vectorial quantities relevant to solvation and VSE shifts exemplified using a CF$_2$ group and its antisymmetric stretching vibration. Left: Dipole moment ($\vec{M}$) and average electric field ($\vec{F}$) of a CF$_2$ group relevant to solvation electrostatics. The average electric field, $\vec{F}$, induced by the solvent follows the direction of the static dipole moment, $\vec{M}$, of the CF$_2$ group as described by Onsager theory. Right: Transition dipole moment ($\overrightarrow{TDM}$) and difference dipole ($\Delta\vec{\mu}$) of $\nu_{as}$(CF$_2$) vibration mode. The $\overrightarrow{TDM}$ is a measure of peak intensity in IR spectra and specifies the direction that maximizes intensity when using polarized light. The difference dipole, $\Delta\vec{\mu}$, describes the electric-field sensitivity of the peak position in vibrational spectra and specifies the direction that maximizes the frequency shift of the mode in an electric field (see eq. 1). The Stark tuning rate is the magnitude of the difference dipole, $|\Delta\vec{\mu}|$.

Despite the growing popularity of fluorination, few studies have explored C-F bonds as VSE probes. This is to a large degree owed to the fact that, despite its strong oscillator strength, the C-F stretching mode, ν(CF), is located in a crowded spectral region of 1300 – 900 cm$^{-1}$ together with C-C stretches and/or C-H deformations.[42] Nevertheless, experimental work by Suydam & Boxer demonstrated that the aromatic ν(CF) of fluorobenzene (FB) shows a strong electric-field sensitivity using vibrational Stark spectroscopy (VSS), a method where a defined external, homogeneous electric field is applied to a VSE probe while measuring its infrared (IR) spectrum.[43] This was further confirmed in computational density functional theory (DFT) studies by Choi & Cho,[42,44] showing a linear electric-field sensitivity of the C-F stretch frequency. In addition, the vibrational response to H-bonding was explored. As observed for the two most popular VSE probes ν(C=O) and ν(C≡N), H-bonding can either induce a linear Stark shift according to eq. 1 or cause a strong deviation from linearity, respectively, via the so-called H-bond blue shift.[19] Theoretical work indicated that the ν(CF) of FB obeys the linear VSE, similar to ν(C=O).[42]

In the course of examining a wider range of fluorinated organic compounds, we were interested in the usefulness of other ν(CF) modes for the determination of electric fields using the VSE. In addition to the abovementioned approaches, a convenient experiment-based method to investigate and, at the same time, calibrate the VSE of such vibrational probes is the combination of the vibrational solvatochromism and molecular dynamics (MD) simulations.[19] In the vibrational solvatochromism, a solute molecule is dissolved in solvents of different polarity and/or H-bonding properties (typically ranging from alkanes to water). This approach exposes the VSE probe to a large range of electrostatic solvation environments and the VSE is detected using IR spectroscopy.[21,22,45,46] The same conditions are reproduced in MD simulations, from which electric fields can be extracted. The correlation of experimental vibrational frequency and simulated electric fields can be modelled by equation 1, which provides the solvatochromic slope, $m$ (units of cm$^{-1}$/ (MV/cm)). This slope describes the sensitivity to the solvent electric field and corresponds to the Stark tuning rate, $|\Delta\vec{\mu}|$.[46] The advantages of this approach are that (a) an absolute calibration is obtained, i.e., the zero-field frequency represents the solute in vacuum; (b) the solute is exposed to molecular environments with a large range of electric fields of up to 100 MV/cm, which are relevant for investigations in more complex settings such as proteins; and (c) the difference between H-bonding and non-H-bonding environments can be assessed using protic and aprotic solvents, respectively.[21] The approach of MD-assisted vibrational solvatochromism is supported by ab initio and QM/MM simulations,[47–49] the solvent-independent anharmonicity of VSE probes like C=O[50] as well as the correspondence to results from direct VSS.[21,46]

Towards exploring the VSE of three basic patterns of aliphatic fluorination, i.e., aliphatic CF, CF$_2$ and CF$_3$ groups, we employed DFT calculations and MD-assisted vibrational solvatochromism, and compared the results to the behavior of aromatic ν(CF) as the best-studied example (see structures in FIG. 2). Using DFT, we estimate the magnitude and the orientation of the vector of the difference dipole underlying the Stark tuning rate for relevant normal modes. With this information, we employed the MD-assisted vibrational solvatochromism using fixed-charge and polarizable MD force fields (AMBER[51,52] and AMOEBA[53], respectively). Fixed charge force fields like AMBER are a standard choice for evaluating VSE tuning rates. While it may seem unnecessary to introduce polarizability when investigating polyfluorinated probes, given the exceptionally low polarizability of C-F bonds,[13] a more precise description can be achieved through the utilization of a multipole expansion. AMOEBA incorporates polarizability and multipoles up to the quadrupole approximation. In our study, we utilize both AMBER and AMOEBA force fields to compare their accuracy and to assess the necessity of a high-level model for calculating VSE tuning rates. From these force fields, we derive the solvatochromic Stark tuning rates for the ν(CF$_n$), extract the influence of conformational flexibility and address the influence of the level of MD theory on solvation electrostatics. The results demonstrate the strengths and limitations of VSE-based analyses of aliphatic C-F vibrations. Our conclusions are of direct relevance for experimental studies of fluorinated molecules using vibrational spectroscopic methods and applicable to a broad variety of existing pharmaceuticals as well as for the study of fluorinated peptides and proteins.

## II. MATERIALS AND METHODS

Materials. As fluorinated probes, 1-fluoro-benzene (FB, Sigma-Aldrich, 99%), 1-fluoro-benzyl alcohol (FBA, Sigma-Aldrich, 99%), 1-fluorohexane (MFH, Synquest labs, 99%), 1,1-difluorocyclohexane (DFcH, abcr 98%) and 1,1,1-trifluorohexane (TFH, Synquest labs, 97%) were used. A collection of 6 aprotic solvents was chosen: n-dodecane (DODEC, Merck, 99%), tetrachloroethylene (TCE, abcr, 99%), trichloromethane (TCM, Carl Roth, 99.0%), toluene (Tol, Carl Roth, 99.5%), acetonitrile (ACN, Sigma-Aldrich, 99.9%) and dimethyl sulfoxide (DMSO, 99.8%, Carl Roth); as well as 3 protic solvents: isopropanol (i-PrOH, 99.8%, Carl Roth), ethanol (EtOH, 99.8, Carl Roth), and water (18.2 MΩ cm$^2$).

Density Functional Theory, Normal Mode Analysis and Spectral Assignment. Computational IR spectra of the fluorinated VSE probes were calculated by normal mode analysis (NMA) after optimization using density functional theory (DFT) under Gaussian 16[54] with the B3LYP functional[55–57] and the 6-311++G(d,p) basis set[58,59]. Spectra are displayed with Gaussian shapes and peak widths of 10 cm$^{-1}$ (FWHM). DFT-based IR spectra for mixtures of conformations in the zero-field approximation (i.e. in vacuum) were calculated as the sum of DFT-IR spectra for all possible conformers of the fluorinated probes, weighted by fractions from MD simulations (AMBER, see below). DFT-calculated VSE difference dipoles were obtained by simulating spectra with electric field values between -30 and +30 MV/cm along different directions, including at least 3 linearly independent ones. The dependency of the vibrational frequency ($\bar{v}$) on the electric field ($\vec{F}$) was approximated to the second degree by considering the difference dipole ($\Delta\vec{\mu}$) and the difference polarizability tensor ($\Delta\underline{\alpha}$) as shown in equation 2. The difference dipole vector (3 parameters - $\mu_x$, $\mu_y$, $\mu_z$), difference polarizability tensor (only diagonal elements - $\alpha_{xx}$, $\alpha_{yy}$, $\alpha_{zz}$) and zero-field frequencies ($\bar{v}_0$) were modelled by least-squares fitting of the frequency values at different fields (see results in TABLE S1). Band assignment was performed via potential energy decomposition (PED) for the two most prevalent conformers of each probe using VEDA (vibrational energy distribution analysis);[60] results using vibrational mode automatic relevance determination (VMARD), a Bayesian regression estimating the most prominent internal coordinates, provided similar results (not shown).[61]

Infrared Spectroscopy and Vibrational Solvatochromism. Experimental IR spectra were recorded using a Bruker 70v spectrometer in attenuated total reflection (ATR) configuration using a single-reflection silicon crystal from IRUBIS[62,63]. Background spectra were recorded with 200 μl of the solvent of choice; solute spectra were obtained from 10 mM solutions in the same solvent. In the case of MFH and TFH, a higher concentration was used (150 mM) due to their lower oscillator strengths. A total of 512 co-additions were acquired for each spectrum at a spectral resolution of 1 cm$^{-1}$.

Fixed-Charge and Polarizable Molecular Dynamics. MD simulations were performed using the fixed-charge general AMBER force-field (GAFF)[51,52] and the multipolar and polarizable amoeba09 force field (atomic multipole optimized energetics for biomolecular simulation, AMOEBA) using GROMACS 2020[64] and TINKER9[65,66], respectively, as described previously.[21,45] The fluorinated solutes were parameterized for AMBER and AMOEBA MD simulations using AmberTools18[67] (AM1-BCC charge model;[68] input structures optimized with B3LYP/6-311++G(d,p)) and Poltype2[69] (electrostatic potential fits to MP2/6-311++G(d,p) calculations), respectively, and are deposited under https://github.com/KozuchLab/Publications/tree/main/VSE_of_CF_probes. Solvent parameters were used from previous work (virtualchemistry.org[70,71] for AMBER; ref. [21,23] for AMOEBA) or as implemented in the force fields. In both cases, a single molecule of a fluorinated probe was simulated in a 4 nm sized cubic box filled with solvent. For AMBER simulations, the system was minimized and equilibrated as NPT ensemble using the Berendsen barostat with a time constant of 1 ps over 200 ps. Production simulations were done using the Parrinello-Rahman pressure coupling, also with a time constant of 1 ps, for a total of 10 ns with van der Waals and Coulomb cut-offs at 1.2 nm and PME for long-range electrostatics. Both, equilibration and production simulations, were done using 2 fs steps and the SD integrator. For simulations in TCE, the MD integrator was used instead to yield a stable configuration. For AMOEBA simulations, minimization was followed by NVT and NPT equilibration over 100 ps at 300 K (1 bar in the latter; induced dipole convergence threshold of 10$^{-2}$ D; mutual polarization; van der Waals and electrostatics cut-offs of 9 Å and 7 Å, respectively, with PME method and van der Waals corrections). The RESPA integrator was used with the Bussi thermostat and molecular volume-scaling, and the Monte-Carlo barostat (time constant of 1 ps) was added during NPT. MD production runs were performed over 10 ns at similar conditions but with van der Waals cut-offs and induced dipole convergence threshold set to 12 Å and $10^{-5}$ D, respectively.

Quantification of Electric Fields. Electric field vectors were calculated on 10 000 equally spaced frames from the electrostatic forces acting on (in AMBER) or from the induced atomic dipoles (in AMOEBA)[21,45] on F atoms and C atoms of the $CF_n$ unit. In both cases, the intramolecular contributions were removed to exclusively obtain the electric field of the solvent by "rerunning" the simulation with similar coordinates but without partial charges on the solute (in AMBER) or in the absence of the solvent (in AMOEBA). Electric fields were recovered from electrostatic forces and induced dipoles by considering the partial charges (AMBER) and polarizabilities (AMOEBA), respectively. The atomistic electric-field vectors were averaged to obtain the effective electric field on the $CF_n$ unit. The time-averaged electric field on the $CF_n$ groups from each solvent was determined by a single Gaussian fit to the histogram of electric field distributions on the x, y and z components (see FIG. S1-S6 and TABLE S2), where z was defined as the main symmetry axis of the $CF_n$ groups (see below).

## III. RESULTS

### A. Assignment of $\nu(CF_n)$-based normal modes using DFT and NMA

Normal modes that contain $\nu(CF_n)$-contributions are found in the complex and crowded spectral region of 1300 – 900 cm$^{-1}$. In order to identify and assign the relevant modes of the aliphatic MFH, DFcH and TFH as well as of the aromatic FB, we compared computed IR spectra obtained from NMA after in vacuo DFT-based optimization with experimental IR spectra in TCE solutions (FIG. 2). TCE was chosen as solvent for this comparison because of its low polarity (dielectric constant $\varepsilon_r \approx 2.5$) and its transparency in the mid-IR window of > 900 cm$^{-1}$. For FB and DFcH only one conformation was relevant (the chair conformation for the latter) in the DFT calculations. Instead MFH and TFH adopt a range of conformations in solution, which can affect peak positions in the IR spectra. To model this conformational heterogeneity, we determined the relative fractions of the relevant conformers from MD simulations of the solutes in TCE. These fractions were used to calculate a weighted average of the DFT-based IR spectra of the individual conformers. Comparing the experimental and computational spectra we note an excellent match at the chosen level of theory (B3LYP/6-311++g(d,p)) with overall consistent relative intensities and peak positions for all four molecules (FIG. 2A-D). Importantly, the DFT-based conformer-averaged IR spectra of MFH and TFH are also in line with the experimental results. The remaining differences relate to the relative intensities at ~1010 cm$^{-1}$ for MFH (FIG. 2B) and ~1099 cm$^{-1}$ for TFH (FIG. 2D) and the spectral shape in the ~1257 cm$^{-1}$ region of TFH (FIG. 2D), as will be discussed below.

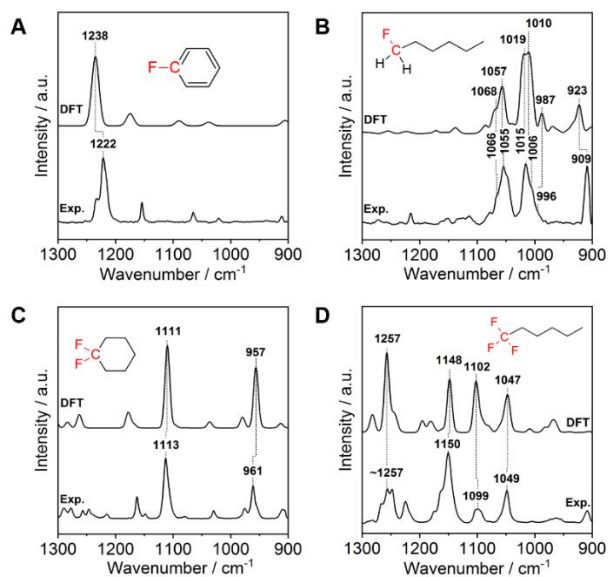

FIG. 2. Experimental and DFT-predicted IR spectra of FB (A), MFH (B), DFcH (C) and TFH (D). Experimental spectra of the molecules were recorded in TCE solution. Theoretical spectra were computed from the weighted sum of the spectra of different conformers using the relative populations estimated by MD simulations in the same solvent. RMSDs of peak positions between experimental and computational data are 18 (FB), 7 (MFH), 10 (DFcH) and 3 (TFH) cm$^{-1}$.

We performed a PED analysis using VEDA[60] to quantify the contribution of the $\nu(CF_n)$ to these normal modes. Starting with FB, the most studied fluorinated compound in terms of the VSE, we note a dominant band at 1222 cm$^{-1}$ in the experimental IR spectrum that corresponds to a peak at 1238 cm$^{-1}$ in the DFT spectrum (FIG. 2A). Consistent with the strong oscillator strength of the $\nu(CF)$, this feature is assigned to a normal mode with $\nu(CF)$ and $\nu(CC)_{ring}$ contributions, and with the highest $\nu(CF)$ character of 45% (TABLE I). Other modes in the analyzed spectral region have a negligible contribution of the $\nu(CF)$, which is consistent with their much smaller intensities.

Despite having only one aliphatic CF bond, the spectrum of MFH is more complicated than that of the aromatic FB showing multiple bands in the 1100 – 900 cm$^{-1}$ region. This complexity is due to two factors: coupling of $\nu(CF)$ and $\nu(CC)$ vibrations leads to the appearance of multiple modes; and conformational heterogeneity can lead to further changes in the spectra. To incorporate both factors, we studied the normal modes of two possible rotamers around the F-C-C-C dihedral, i.e., trans (t) and the gauche conformations (1-gauche, g) with an otherwise extended structure (FIG. 3B.1). Each of the conformers exhibits three dominant $\nu(CF) + \nu(CC)$ modes with varying symmetry and with $\nu(CF)$ contribution between 16 – 40% (TABLE I). In particular, the bands at 1015 and 996 cm$^{-1}$ in the experimental spectra, which are assigned to the trans conformer (at 1019 and 987 cm$^{-1}$ in DFT-based spectra), have the highest ν(CF) character with 40% and 41%, respectively, but are located in a crowded region overlapping with bands of the gauche conformer. The latter are the 1055 and 1006 cm$^{-1}$ modes (1057 and 1010 cm$^{-1}$ in DFT-based spectra). The only band unaffected by spectral overlap is found at 909 cm$^{-1}$ (923 cm$^{-1}$ in DFT-based spectra), which originates exclusively from the gauche rotamer of the F-C-C-C dihedral with 20% ν(CF) character.

DFcH exists in solution in the chair conformation. This reduces the complexity of the IR spectrum, such that a very good correspondence between experimental and computed spectra is observed with prominent peaks at 1113 and 961 cm$^{-1}$ (1111 and 957 cm$^{-1}$, respectively, in DFT-based spectra; FIG. 2C). Based on the CF$_2$ motif, a symmetric ν$_s$(CF$_2$) and antisymmetric ν$_{as}$(CF$_2$) mode would be expected, which is indeed the case in the twisted boat conformation, where both C-F bonds are equivalent (see FIG. 3C.1). However, the chair structure places one C-F in axial and one in equatorial position, resulting in an assignment of the 1113 and 961 cm$^{-1}$ bands to normal modes of ν$_{eq}$(CF) + ν(CC) and ν$_{ax}$(CF) + ν(CC) and considerable ν(CF) character of > 42 % (TABLE I).

Despite the presence of different conformers in TFH, its spectrum is less complicated since all rotamers of the F-C-C-C dihedral are equivalent and instead only more distant dihedrals influence the spectra. Again, we considered the trans and gauche rotamers at the most adjacent C-C-C-C dihedral (all other dihedrals are extended, see FIG. 3D.1), which are the two most dominant structural features during MD simulations. We note four features at ~1257, 1150, 1099, and 1049 cm$^{-1}$ in the experimental spectra with excellent correspondence to peaks at 1257, 1148, 1102, and 1047 cm$^{-1}$ (FIG. 2D) in the conformer-averaged DFT spectra with ν(CF) character of 19 – 48% (TABLE I). Here, the expected symmetric ν$_s$(CF$_3$) and antisymmetric ν$_{as}$(CF$_3$) modes are largely maintained but are coupled to C-C-C deformations, δ(CCC), yielding one ν$_s$(CF$_3$) + δ(CCC) and three ν$_{as}$(CF$_3$) + δ(CCC) modes for the high-frequency mode and the three lower frequency modes, respectively.

TABLE I. Band assignment from DFT normal mode analysis and peaks with highest C-F stretch contribution after PED decomposition.

| Exp. $\bar{\nu}$ / cm$^{-1}$ | DFT $\bar{\nu}$ / cm$^{-1}$ | Assignment | ν(CF) content (PED) |
|---|---|---|---|
| **FB** | | | |
| 1222 | 1238 | ν(CF) + ν(CC)$_{ring}$ | 45% |
| **MFH** | | | |
| 1066 | 1068 (t/g) | ν(CF) + ν(CC) | 40% |
| 1055 | 1057 (g) | ν(CF) + ν(CC) | 19% |
| 1015 | 1019 (t) | ν(CF) + ν(CC) | 40% |
| 1006 | 1010 (g) | ν(CF) + ν(CC) | 16% |
| 996 | 987 (t) | ν(CF) + ν(CC) | 41% |
| 909 | 923 (g) | ν(CF) + ν(CC) | 20% |
| **DFcH** | | | |
| 1113 | 1111 | ν$_{eq}$(CF) + δ(CCC) | 42% |
| 961 | 957 | ν$_{ax}$(CF) + δ(CCC) | 45% |
| **TFH** | | | |
| 1257 | 1257 (t/g) | ν$_s$(CF$_3$) + δ(CCC) | 48% |
| 1150 | 1148 (t/g) | ν$_{as}$(CF$_3$) + δ(CCC) | 19% |
| 1099 | 1102 (t/g) | ν$_{as}$(CF$_3$) + δ(CCC) | 30% |
| 1049 | 1047 (t) | ν$_{as}$(CF$_3$) + δ(CCC) | 32% |

ν=stretching, δ=deformation, eq=equatorial, ax=axial, as=antisymmetric, s=symmetric

### B. Difference dipole magnitudes (Stark tuning rates) and orientations from DFT

To determine the magnitudes and orientations of the difference dipoles of the modes involving $\nu(CF)$, as assigned above, vibrational frequencies were computed in the presence of electric fields of varying strengths ranging from +30 to -30 MV/cm, and of (at least three) different orientations with respect to the main axis of the $CF_n$ groups. The negative sign of the electric fields denotes a stabilizing interaction with the C-F dipole, as experienced in a solvating environment.[19] Applying eq. 1 allowed us to accurately model all sampled frequencies obtained from DFT calculations with electric fields along different directions.

As previously reported,[42] we observe a considerable, monotonous redshift of the aromatic $\nu(CF)$ peak at 1235 cm$^{-1}$ by ca. -25 cm$^{-1}$ with increasing (negative) electric fields along the C-F axis of FB (FIG. 3A), while off-axis fields did not result in any relevant shifts. Due to the resulting linear trend between vibrational frequency and electric field (FIG. 3A.3), quadratic terms of eq. 1 were negligible (TABLE S1) and a difference dipole of 0.63 cm$^{-1}$/(MV/cm) oriented along the C-F bond was determined; a value similar to the DFT-based Stark tuning rates determined previously[42].

As in the spectral assignment above, we performed the DFT analysis for the aliphatic $\nu(CF)$ modes of MFH on the two relevant conformations of the terminal -CH$_2$F group (i.e. trans and gauche, FIG. 3B.1, all other torsions were kept as trans) as a proxy for all-trans and gauche rotamers. In the trans conformation, the two $\nu(CF)$ modes at ca. 1023 and 988 cm$^{-1}$ were most sensitive to the electric field (FIG. 3B.2, top), which we denote t1 and t2. Increasingly negative electric fields along the C-F axis (from 0 to -30 MV/cm) caused a redshift by -13 and -23 cm$^{-1}$ for t1 and t2, respectively, which correspond to Stark tuning rates of 0.48 and 0.51 cm$^{-1}$/(MV/cm) (FIG. 3B.3, top). The associated vectors were largely aligned along the C-F axis (FIG. 3,B.1) with slight tilts by 8.7° and 17.6° in the F-C-C plane. In the gauche conformation, the modes at 1002 and 923 cm$^{-1}$ (g1 and g2, respectively) are electric field sensitive and shift by ca. -8 cm$^{-1}$ in electric fields of -30 MV/cm (FIG. 3B.2, bottom). This results in Stark tuning rates of 0.23 and 0.26 cm$^{-1}$/(MV/cm) (FIG. 3B.3, bottom), respectively, whose vectors are considerably rotated by 25.6° and 39.7° (FIG. 3B.1) around the C-CH$_2$F bond as a result of the mixed $\nu(CF) + \nu(CC)$ character of the modes.

Even though the twisted boat is not a relevant conformation of DFcH in solution, the equivalence of both C-F bonds (FIG. 3C.1, left) makes it an interesting conceptual reference system for the VSE of the CF$_2$ group. The $\nu_s(CF) + \delta(CCC)$ and $\nu_{as}(CF) + \delta(CCC)$ modes are found at 1100 and 940 cm$^{-1}$ (tb1 and tb2) and shift by -10 to -15 cm$^{-1}$ in fields of -30 MV/cm along the CF$_2$ bisector (FIG. 3C.2, top). The associated difference dipole magnitudes are 0.26 and 0.41 cm$^{-1}$/(MV/cm) and, despite being of orthogonal CF$_2$ displacement geometry and transition dipole orientation, both are directed along the bisector (FIG. 3C.1, left). While this is counterintuitive at first glance, it is in line with the underlying anharmonicities of the two C-F bonds (i.e. both C-F bonds elongate upon vibrational excitation). The change in the normal mode composition in the chair conformation to $\nu_{eq}(CF) + \nu(CC)$ and $\nu_{ax}(CF) + \nu(CC)$ leads to a shift of the modes to 1110 and 957 cm$^{-1}$ (c1 and c2), which redshift by similar extend in fields along the bisector (FIG. 3C.2, bottom). However, the difference dipoles changed such that they are directed along the CF$_{eq}$ or between bisector and CF$_{ax}$, respectively, with magnitudes of 0.72 and 0.42 cm$^{-1}$/(MV/cm) (FIG. 3C.1, right; FIG. 3C.3, bottom).

The most sensitive modes of TFH in both trans and gauche rotamers were the bands at ~1255 cm$^{-1}$ (t1 and g1) and at 1148 cm$^{-1}$ (t2 and g2) (FIG 3D.1, 3D.2), which redshifted by -8 to -10 cm$^{-1}$ in fields along the C3 axis of the CF$_3$ group. Based on the contribution of the $\delta(CCC)$, however, the difference dipoles of the two modes are oriented along one C-F bond or along the bisector of a CF$_2$ (FIG 3D.1). This leads to Stark tuning rates of (t1) 0.21 and (t2) 0.49 cm$^{-1}$/(MV/cm) or (g1) 0.33 and (g2) 0.29 cm$^{-1}$/(MV/cm) (FIG. 3D.3). As discussed later, the fluctuating orientation of the difference dipoles is a likely origin for the intricate band shape at around ~1257 cm$^{-1}$.

### C. Molecular dynamics-assisted vibrational solvatochromism

As an experimental equivalent to the DFT-based IR spectra in external electric fields (FIG. 3), we recorded IR spectra of the vibrational solvatochromism to assess the corresponding peak shifts in various electrostatic solvation environments. For this, we used aprotic solvents of increasing polarity from DODEC to DMSO (see specified solvents in FIG. 5 and 6) as well as the protic solvents i-PrOH, EtOH, and water. We report the spectra for solute/solvent mixtures where sufficient solubility was obtained and solute bands where clearly detectable. FB had residual solubility in water, whereas the aliphatic compounds were not sufficiently soluble. To detect spectra in H-bonding situations, we aimed at hydroxylated derivatives, which are reported below for FB; aliphatic alcohols showed considerable overlap with the $\nu(C-O)$ impeding the analysis and are therefore not discussed herein. Average solvent electric fields were determined from MD simulations using the fixed-charge AMBER[51,52] and the multipolar and polarizable AMOEBA force field,[53] to assess the benefit of the more accurate electrostatic description of the latter.

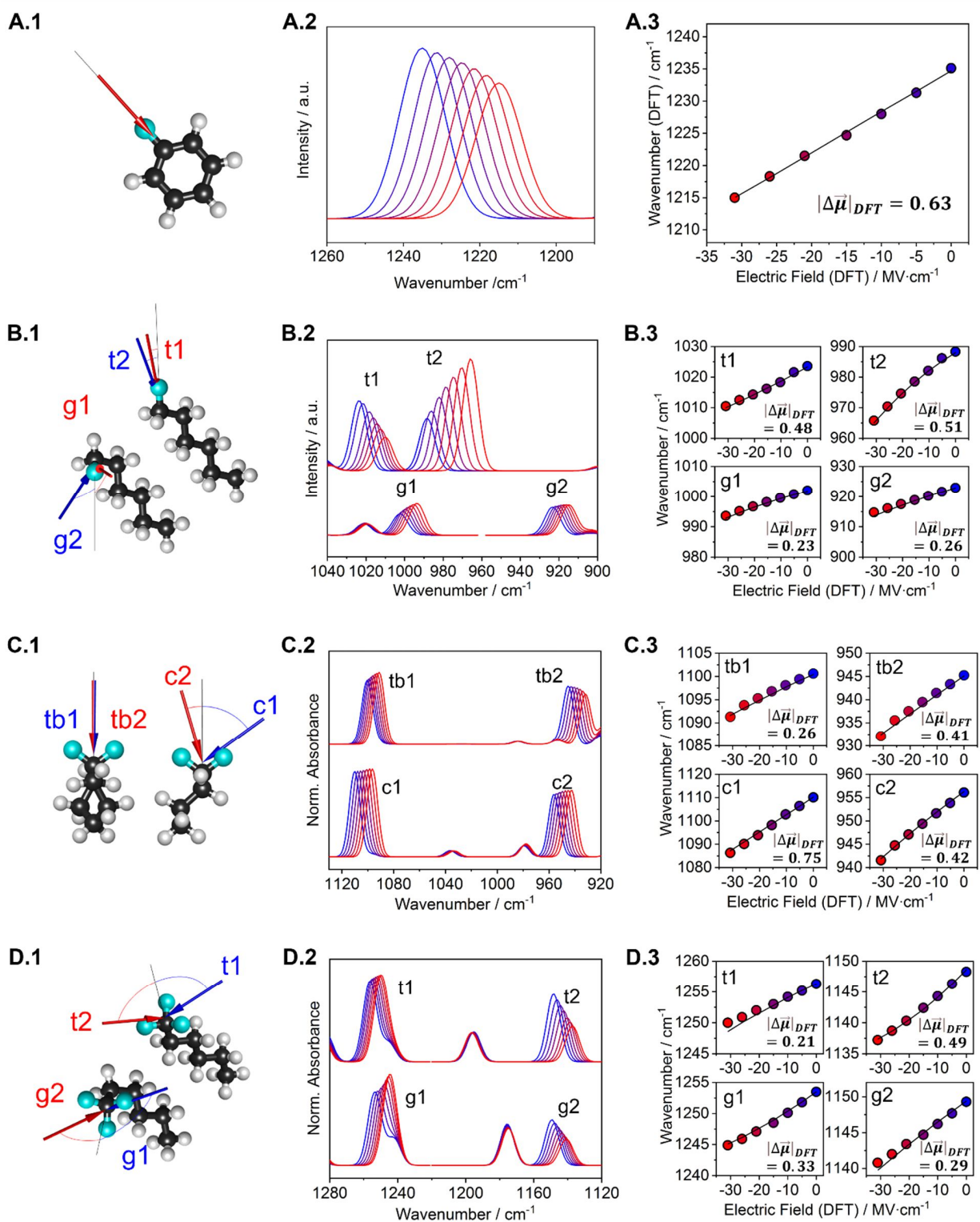

FIG. 3. Vibrational Stark shifts of fluorinated probes from DFT calculations under homogenous fields for (A) FB, (B) MFH, (C) DFcH and (D) TFH. First column: Molecular structures with calculated difference dipole direction for FB (A.1); extended (top) and 1-gauche (bottom) conformers of MFH (B.1); twisted-boat (left) and chair (right) conformers of DFcH (C.1); and all-trans (top) and 1-gauche (bottom) conformers of TFH (D.1). The labels t1/t2, g1/g2, tb1/tb2 and c1/c2 refer to ν(CF)-containing modes of trans, gauche, twisted boat and chair conformation; see main text for details. Second column: Peak positions of C-F stretching modes for these conformers with electric fields between 0 (blue) and -30 MV/cm (red). Third column: Peak position correlation with electric field values and estimated VSE tuning rates for electric fields along the difference dipole vector direction. Black line shows the frequencies predicted by the difference dipole and diagonal polarizability model. Accuracy of $R^2 = 0.99$; see TABLE S1 for all results.

As implied by the potentially varying orientations of the vectorial quantities in FIG. 1, it is imperative to determine the orientation of the solvent electric field onto the VSE probe for a correct application of our solvatochromic calibrations. We defined the main symmetry axes of the $CF_n$ groups and the direction towards the adjacent C atom as the z and x axes, respectively (see insets in FIG. 4A, B, C, D or FIG. S1 for frame definitions). Using this frame of reference, we quantified the projections of the electric field strengths onto the x, y and z axes. As shown for the example of the solvent field of DMSO (FIG. 4), the solvent field was oriented on average largely along the main symmetry axes of the VSE probes, i.e. the C-F bond, the $CF_2$ bisector or the C3 axis of the $CF_3$ group for mono-, di- and trifluorinated compounds, respectively (see FIG. S2-S6 for other solvents). A maximum deviation of the electric field vector by 7.3° was observed towards the x-axis in MFH with AMBER force field, which accounted for 13 % of the total solvent electric field magnitude. Average deviation for all probes were 2.5° and 1.9°, in AMBER and AMOEBA force-fields respectively. As such, we can use the electric field magnitudes along the z-axis in the following solvatochromic frequency/field correlations. We denote the solvatochromic slopes obtained in this section as $m$ to distinguish between the Stark tuning rates in defined external fields.

Fluorobenzene. Analogously to the DFT-predicted IR spectra, we observe that the band at 1220 cm$^{-1}$ of FB (FIG. 5 A.1) experiences a considerable redshift by -10 to -15 cm$^{-1}$ from DODEC to DMSO, i.e., towards more polar solvents. In some solvents, a shoulder at 1233 cm$^{-1}$ can be detected, which however does not experience any solvatochromic shift. We assigned this peak tentatively to dimeric species as suggested also for other Stark probes in solvatochromic experiments.[24,46] Against intuition, FB in water does not show the most redshifted peak, despite H-bonding typically exerting electric fields of the highest magnitude,[21,22,45,49,72,73] but appears at a similar position as in toluene. A possible scenario to explain this observation is that H-bonding is avoided by the hydrophobic FB, as observed for many fluorinated molecules.[74] To test this hypothesis, we recorded solvatochromic IR spectra of 4-fluorobenzyl alcohol (FBA), which has a higher solubility in water and also shows a ν(CF) mode at ~1220 cm$^{-1}$ that redshifted vs. DMSO as solvent (FIG. 5A.2). Interestingly, the spectrum in water shows an asymmetric band shape with two components at ~1220 and 1209 cm$^{-1}$, which can be tentatively assigned to a non-H-bonded and H-bonded fraction, respectively. Extracting the MD-based electric fields, we observed that the electric fields from both force fields varied considerably (FIG. 5A.3): whereas AMBER electric fields on FB range from 0 to ca. -10 MV/cm, AMOEBA predicted averaged solvation fields were consistently higher by a factor of ~2. We can rationalize this observation based on the quality of the electrostatic parameters of the force fields (see Discussion). Correlating vibrational frequencies and fields for FB (FIG. 5A.3), we observe a linear trend with an AMOEBA-based solvatochromic tuning rate of $m$ = 0.38 cm$^{-1}$/(MV/cm). For AMBER fields, the slope is 0.8 cm$^{-1}$/(MV/cm). Overall, the MD simulations (and the extracted fields) were in line with a situation where FB avoids direct H-bonding via its C-F bond resulting in lowered fields. However, the distributions of electric fields of FBA in water showed an additional high-field shoulder (at -12 or -42 MV/cm for AMBER and AMOEBA, respectively) due to a fraction of H-bonded FBA molecules (FIG. 5A.1). The data point of the H-bonded FBA fraction falls well on the solvatochromic trend with considerably higher electric fields than DMSO, which is in line with linear VSE behavior in aprotic and protic solvents as predicted by previous work.[42]

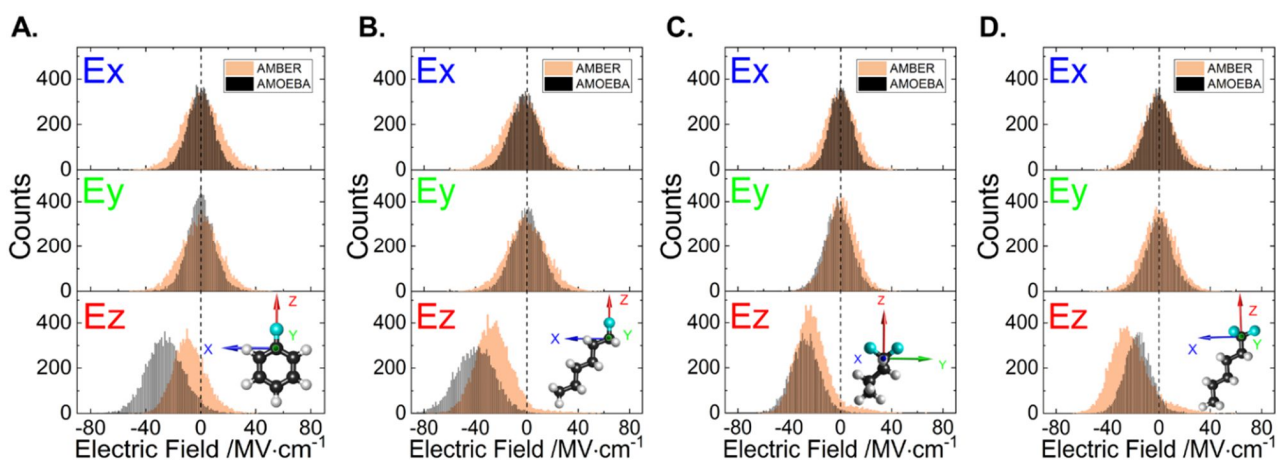

FIG. 4. Histogram of electric field values from MD simulations in DMSO for (A) FB, (B) MFH, (C) DFcH and (D) TFH along x,y and z directions. Cartesian coordinates defined using the local reference frame shown in the molecular models. Electric field distributions for other solvents are shown in FIG. S1-S6.

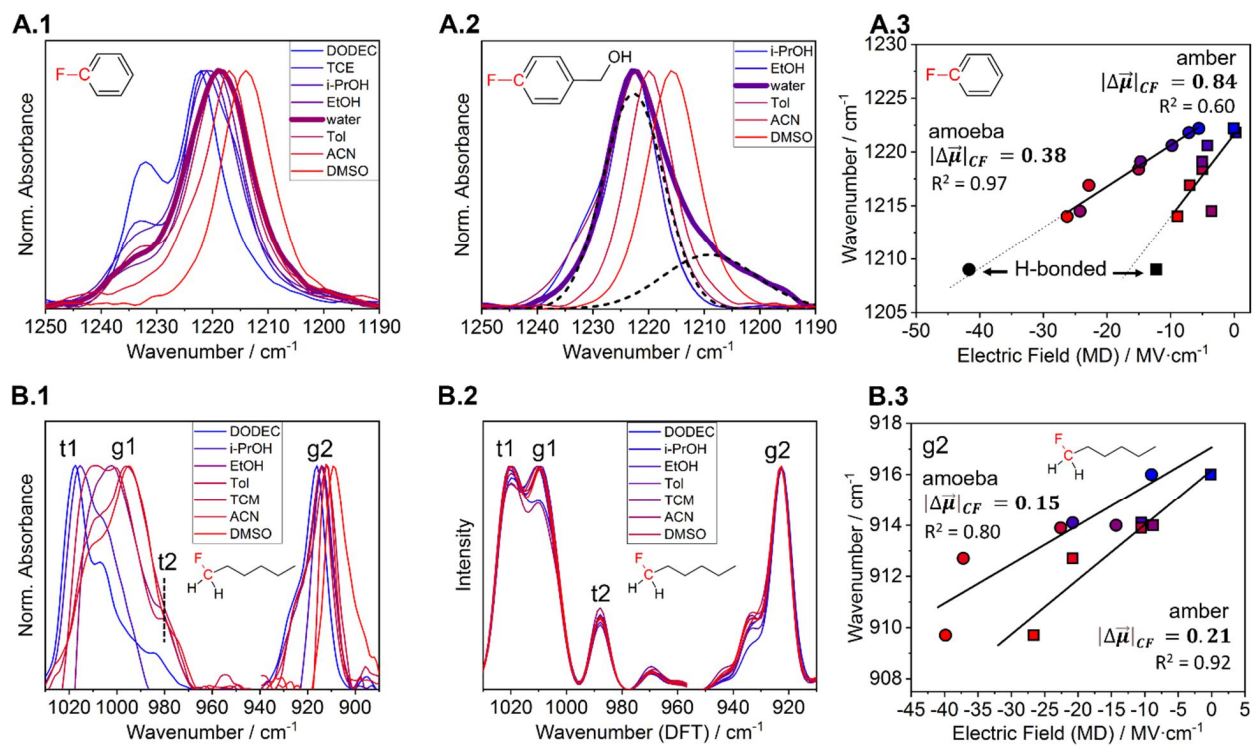

FIG. 5. MD-assisted solvatochromic field-frequency calibration of FB and MFH. **A.1**, **A.2**, **B.1**: Experimental IR spectra of monofluorinated probes, the aromatic FB (A.1), its water-soluble variant FBA (A.2) and the aliphatic MFH (B.1). **B.2**: DFT spectra of MFH at zero electric field from the conformational mixtures on each solvent estimated from MD simulations. **A.3, B.3,**: Correlation of experimental peak positions and electric fields from AMBER and AMOEBA MD simulations for FB (A.3) and MFH (B.3). Hydrogen-bonded values in A.3 obtained from water-soluble FBA. See all fitting results in TABLE S2.

1-(Mono)fluorohexane. The spectral region of 1050–900 cm$^{-1}$ of MFH showed the most pronounced and clearest peak shifts in the solvents (FIG. 5B.1) by -30 and -8 MV/cm for the features at ~1000 (overlap of t1, g1, t2) and 920 cm$^{-1}$ (g2), respectively. Due to the conformational flexibility of MFH, it is important to parse these shifts into contributions due to conformational and electrostatic changes across the solvents. Our strategy relied on calculating DFT-based IR spectra of each possible conformation in vacuo, determining the fraction of these conformations in the solvent set from MD simulations and, by this, assembling fraction-weighted average spectra showing the drift of conformational composition only (that is, without solvatochromic contributions to peak shifts; FIG. 5B.2). When classifying all possible conformations based on the trans and gauche rotamer of the terminal -CH$_2$F group, we note that the trans:gauche-ratio shifts from 44:56 to 35:65 from DODEC to DMSO. We notice that the peaks of t1 and g1 show considerable overlap at peak widths of 10 cm$^{-1}$ (FWHM) in the corresponding DFT-based IR spectra, which was chosen to be consistent with the experimental IR widths. In contrast, the modes t2 and g2 are more isolated. Interestingly, while t2 and g2 do not shift throughout the simulated spectra, which is consistent with the absence of environmental electric fields (i.e. DFT-IR spectra obtained in vacuo), the peak shape involving the peaks t1 and g1 experiences observable changes. This is in line with spectral changes due to conformational effects. Turning back to the experimental spectra (FIG. 5B.1), we can conclude that the spectral region including the overlapping t1, g1, and t2 bands convolute solvatochromic and conformational changes in the spectra, whereas g2 gives access to evaluate the electrostatic, solvatochromic shift in the gauche conformation.

Electric fields along the C-F bond of MFH were similar within ±2 MV/cm for different conformations and showed a range of 0 to -30 MV/cm and -10 to -40 MV/cm for AMBER and AMOEBA force fields, respectively. The latter is in line with previous results[21,45] where AMOEBA showed an offset by roughly -10 MV/cm rather than a difference by a considerable factor as determined for FB. Correlating the peak position with MD electric fields, we obtain consistent linear trends with solvatochromic tuning rates along the CF axis of 0.21 and 0.15 cm$^{-1}$/(MV/cm) for AMBER and AMOEBA, respectively (FIG. 5B.3). Due considerable overlap of t1, g1, and t2, we performed a fit to the complicated absorption feature using Gaussian line shapes, which we show in FIG. S7 and S8. Despite the considerable overlap we obtain very good linear field/frequency correlations with AMBER and AMOEBA fields (R$^2$ of > 0.82 and > 0.97, respectively) suggesting

AMOEBA-based solvatochromic tuning rates of 0.27, 0.50, and 0.27 cm$^{-1}$/(MV/cm) for t1, g1, and g2 (0.3, 0.56, and 0.31 cm$^{-1}$/(MV/cm) for AMBER). However, we would like to remind that these values may contain effects due to the conformational drift.

*1,1-Difluorocyclohexane.* Experimental IR spectra of DFcH in different solvents show bands peaking at 1110 (c1) and 960 cm$^{-1}$ (c2), which redshift by -10 or -8 cm$^{-1}$ from DODEC to DMSO in agreement with DFT (FIG. 6A.1). Despite the two distinct positions of the F atoms in either equatorial or axial position, the MD-based electric fields varied by ±2 MV/cm at most among all solvents, which is consistent with a solvent electric field along the bisector (FIG. 6A; see also FIG. S5). Correlating the peak positions with the electric fields along the CF$_2$ bisector from MD simulations, we obtain solvatochromic slopes of 0.43 or 0.40 cm$^{-1}$/(MV/cm) for c1 and 0.42 or 0.3 cm$^{-1}$/(MV/cm) for c2 (both for AMOEBA and AMBER, respectively; FIG. 6C.2, 6C.3). These results can be considered as consistent taking into account the scatter of the data (R$^2$ = 0.71 – 0.98).

*1,1,1-Trifluorohexane.* FIG. 6B.1 shows the spectral region above 1120 cm$^{-1}$, since bands below this range exhibit only minor peak shifts (< 2 cm$^{-1}$) and/or partially erratic behavior. We note a monotonous redshift of the t2/g2 band at 1150 cm$^{-1}$ by -5 cm$^{-1}$ towards polar solvents, whereas the collection of bands at ~1257 cm$^{-1}$ (t1,g1) show a change in relative intensity with minor shifts by ~2 cm$^{-1}$. As for MFH, we reconstructed DFT-based IR spectra for each solvent: taking conformational fractions from MD simulations and corresponding in vacuo computed spectra to estimate the impact of changes in the conformational distribution of TFH to the experimental IR spectra (FIG. 6B.2). Overall, in DODEC a ratio of 92:8 between trans and gauche F$_3$C-C-C-C rotamers was observed in MD simulations, which shifted to 85:15 in DMSO, but no changes were observed for the t1/g1 and t2/g2 bands in the DFT-based spectra. As such, we can assign the experimental shift of the 1150 cm$^{-1}$ to a VSE. Using MD simulation, we find that all three C-F bonds are exposed to similar electric fields (with differences < 2 MV/cm), which is consistent with a solvent electric field directed along the CF$_3$'s main symmetry axis (FIG. 4D; see also FIG. S7). In AMOEBA MD simulations these fields varied between -5 and -15 MV/cm, whereas AMBER force fields showed a span of 0 to -25 MV/cm, resulting in force field-dependent difference by a factor of almost ~2. As for FB, we can rationalize these differences by the quality of electrostatic parameters (see Discussion below). Correlating peak positions of the t2/g2 band with the corresponding electric fields, we obtain relevant linear correlations (R$^2$ ≈ 0.7) that yielded solvatochromic slopes, of 0.16 and 0.34 cm$^{-1}$/(MV/cm), in AMBER and AMOEBA, respectively (FIG. 6B.3).

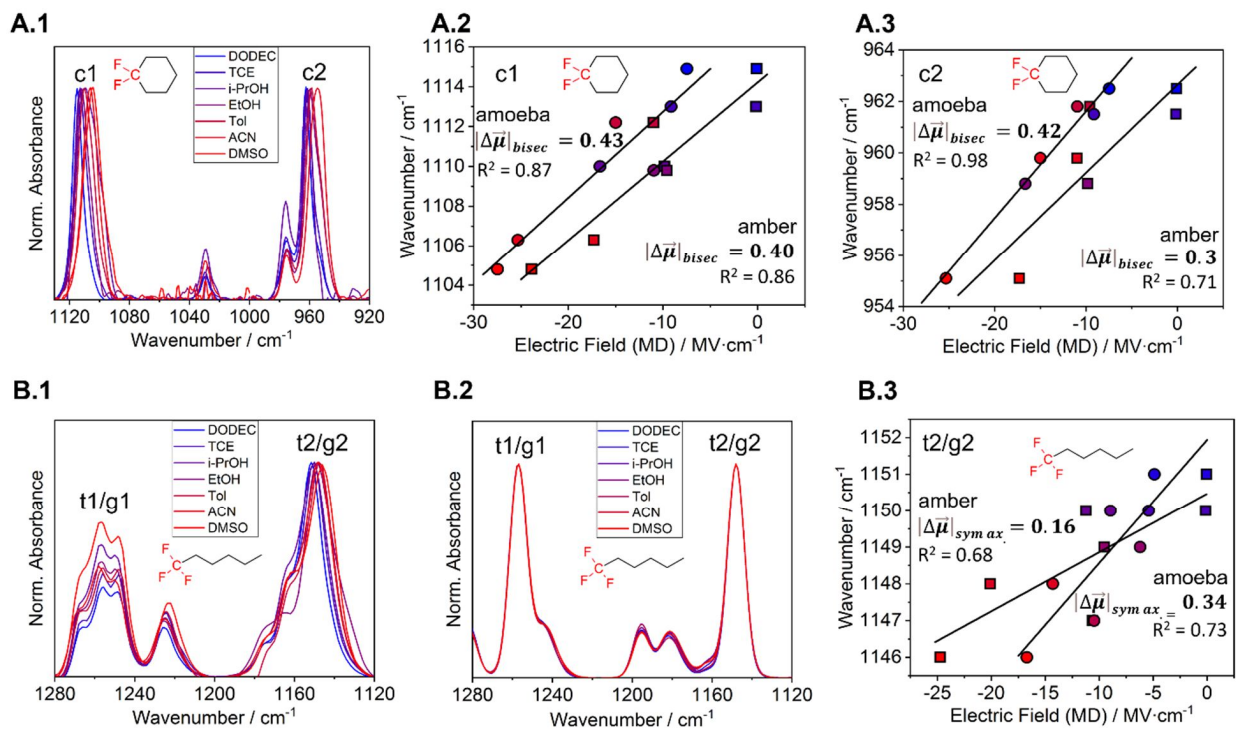

FIG. 6. MD-assisted solvatochromic field-frequency calibration of DFcH and TFH. **A.1**, **B.1**: Experimental IR spectra of polyfluorinated probes DFcH (A.1) and TFH (B.1) in different solvents. **B.2**: DFT spectra of TFH at zero electric field from the conformational mixtures on each solvent estimated by AMBER. **A2**, **A.3**, **B.3**: Correlation of experimental peak positions and electric fields from AMBER and AMOEBA MD simulations for DFcH (A.2, A.3) and TFH (B.3). See all fitting results in TABLE S2.

## IV. DISCUSSION

### A. Stark tuning rates, difference dipoles, solvatochromic slopes, and electric field orientations in solution.

The ability to extract the electrostatic interactions of a molecular environment on a quantitative basis is highly dependent on the accuracy of the available Stark parameters. The results from our work are compiled together with literature data in TABLE II and FIG. 7. In order to provide a meaningful comparison of the Stark tuning rates of $CF_n$ probes in TABLE I, it is important to preface our discussion with considerations of specific differences in the methodologies used to determine Stark parameters. To the best of our knowledge, only VSS and DFT-based analyses have been published on FB and 2,2,2-trifluoro ethanol, the latter containing a $CF_3$ probe. In both approaches, the VSE probe is exposed to external homogeneous electric fields, while spectra have been recorded or vibrational parameters calculated. In VSS, the Stark tuning rate is determined as $f \cdot |\Delta \vec{\mu}|$. Here, $f$ is the local field correction factor, a solvent-independent factor of $2.0 \pm 0.5$, which describes the enhancement of the effective local electric field acting on the VSE probe with respect to the known external electric field.[75] This factor provides a direct link to the MD-assisted vibrational solvatochromism, where we use the local electric fields due to the molecular environment created by the solvents, i.e., an electrostatic environment with $f = 1$. VSS has been applied to FB to yield Stark tuning rates of $0.84/f$ cm$^{-1}$/(MV/cm), but solvatochromic analyses are missing.[43,76] In vacuo DFT-based approaches, as used herein, $f$ is not relevant since molecular or dielectric environment is not included; $|\Delta \vec{\mu}|$ has been determined directly and resulted in 0.63 cm$^{-1}$/(MV/cm) in previous analyses of FB.[44] Taking into account that $f \approx 2$, $|\Delta \vec{\mu}|$ would have been expected to yield a value of 0.4 cm$^{-1}$/(MV/cm). It can be inferred that the DFT-based analysis overestimated the value by ~1.5. As a matter of fact DFT-based tuning rates are overestimated by factors of $1 - 1.5$[21,27,46,77] as a result of various factors including level of theory, missing/inaccurate consideration of anharmonicity, assumptions on the direction of difference dipole vectors, etc. (see discussion in SI of ref. [19]).

In the present work, we have determined a DFT-based Stark tuning rate of FB of 0.63 cm$^{-1}$/(MV/cm), which is in excellent agreement with analyses in literature that used a direct correlation between external field and resulting frequency.[44] As expected from the symmetry of FB, the difference dipole vector is aligned along the C-F bond and the difference polarizability was not necessary to model the peak shifts (FIG. 7A; TABLE S1). This is interesting so far that the 1222 cm$^{-1}$ band contains 45% v(CF) character and thus considerable contribution of the v(CC) vibration of the polarizable benzene ring (TABLE I). In comparison, v(C=O) and v(C≡N) are almost exclusively localized on the C=O or C≡N bonds. The simple linear description of FB's ν(CF) mode, not requiring quadratic terms, points to a picture where the C-F dipole dominates $|\Delta\vec{\mu}|$, despite a considerable contribution of ν(CC) to the normal mode. This is further supported by our solvatochromic analysis, which yielded a slope of 0.38 cm$^{-1}$/(MV/cm) using AMOEBA, consistent with $|\Delta\vec{\mu}| = 0.84/f$ cm$^{-1}$/(MV/cm) and $f \approx 2$, by only using electric fields on the C-F bond and no other atoms of FB. The AMBER-based result, which did not match this relation based on $f \approx 2$, will be discussed below. The correspondence between VSS, DFT and solvatochromism-based results presents a valuable reference point for the discussion of the aliphatic compounds, for which literature data (except for the CF$_3$ group) are not available.

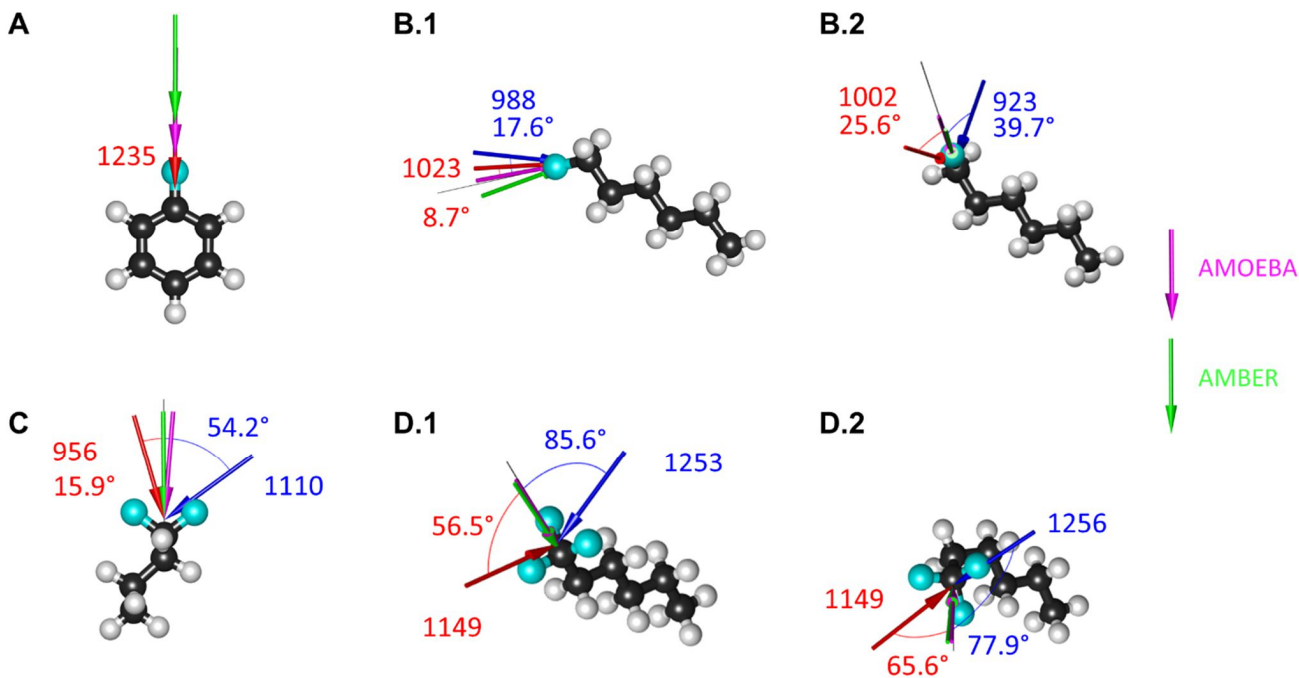

FIG. 7. Comparison of VSE difference dipole vectors (red and blue arrows) for different conformers of FB (A), MFH (B.1 and B.2), DFcH (C), and TFH (D.1 and D.2) with averaged MD electric field directions from AMOEBA and AMBER MD simulations (magenta and green arrows, respectively). Difference dipole vectors are labeled with vibrational frequency and the tilt angles in similar colors. All angles specified are with respect to the main symmetry axes, i.e. the C-F bond for FB and MFH (A, B.1, B.2), the $CF_2$ bisector of DFcH (C), or the $C_3$ axis of the $CF_3$ group (D.1, D.2).

TABLE II. Summary of Stark tuning rates from DFT and solvatochromism with values previously reported in literature. DFT and VSS based data is taken from refs. [44] and [43]. Symmetry axes are shown in FIG. 4 and FIG. S1. Difference dipole vectors determined in this work are shown in FIG. 7; solvatochromic Stark tuning rates are determined according to eq. 2 as discussed in the text.

| Solute | Symmetry axis | Exp. $\bar{\nu}$ / cm$^{-1}$ | Literature | | This work – Stark tuning rates | | This work – solvatochromic slopes | |
|---|---|---|---|---|---|---|---|---|
| | | | $\|\Delta\vec{\mu}\|$ DFT | $f\|\Delta\vec{\mu}\|$ VSS | $\|\Delta\vec{\mu}\|$ DFT | Angle (°) to symm. axis | m – Vib. Solv. AMOEBA wrt. symm. axis (wrt. $\Delta\vec{\mu}$ direction) | m – Vib. Solv. AMBER wrt. symm. axis (wrt. $\Delta\vec{\mu}$ direction) |
| Fluorobenzene (FB) | C-F | 1222 | 0.63 | 0.84 | 0.63 | 0 | 0.38 | 0.8 |
| 1-(Mono)fluoro hexane (MFH) | C-F | 1015 (t1) | - | - | 0.48 | 9 | [0.27 (0.27)]$^a$ | [0.30 (0.30)]$^a$ |
| | | 1006 (g1) | - | - | 0.51 | 26 | [0.50 (0.56)]$^a$ | [0.56 (0.62)]$^a$ |
| | | 996 (t2) | - | - | 0.23 | 18 | [0.27 (0.28)]$^a$ | [0.31 (0.32)]$^a$ |
| | | 909 (g2) | - | - | 0.26 | 40 | 0.15 (0.20) | 0.21 (0.27) |
| 1,1-Difluoro cyclohexane (DFcH) | CF$_2$ axis | 1113 (c1) | - | - | 0.75 | 54 | 0.43 (0.73) | 0.40 (0.68) |
| | | 961 (c2) | - | - | 0.42 | 16 | 0.42 (0.44) | 0.30 (0.44) |
| 1,1,1-Trifluoro hexane (TFH) | CF$_3$ axis | ~1257 (t1) | - | 0.77 | 0.21 | 78 | - | - |
| | | ~1257 (g1) | - | 0.77 | 0.33 | 87 | - | - |
| | | 1150 (t2) | - | - | 0.49 | 66 | 0.34 (0.82) | 0.16 (0.39) |
| | | 1150 (g2) | - | - | 0.29 | 57 | 0.34 (0.61) | 0.16 (0.29) |

$^a$ Values of solvatochromic tuning rates are obtained from overlapping bands that are deconvoluted based on Gaussian peak fits. See FIG. S7 and S8.

In addition to the intricate composition of the normal modes (TABLE I), the aliphatic CF$_n$ groups are further complicated by conformational changes of the carbon longchain impacting the vibrational spectrum. Of the tested molecules, the difluorinated DFcH presents the most straightforward case as it is not affected by conformational heterogeneity due to the cyclic structure. Using DFT, we determined for the ν$_{eq}$(CF) and ν$_{ax}$(CF)-based modes (~1100 and ~960 cm$^{-1}$) considerable Stark tuning rates of 0.75 and 0.42 cm$^{-1}$/(MV/cm), respectively. Their difference dipole vectors were aligned with the equatorial C-F bond or between the CF$_2$ bisector and the axial C-F bond (54° or 16°), respectively (FIG. 7C) due to non-identical character of the C-F bonds. This difference dipole orientation considerably impacts the determination of electric fields in solution. Evaluating the MD-based electrostatics, we find that electric fields exerted by the solvents stabilize the CF$_2$ dipole on average and not the individual C-F dipoles, irrespective of solvent and force fields (FIG. 4C; FIG. S1-S6). This is evident from Gaussian-shaped distributions along the main symmetry axis (here z-axis) around a non-zero field, but zero-centered distributions along the orthogonal axes (x and y). As such, the directly obtained solvatochromic slopes [0.43 and 0.42 cm$^{-1}$/(MV/cm)] relate to the projection of the averaged electric field vector on the difference dipole vectors (i.e. $\vec{F}$ and $\Delta\vec{\mu}$, respectively, in FIG. 1). Considering this situation, we can obtain the Stark tuning rate magnitude from the solvatochromic slopes via

$$m = |\Delta\vec{\mu}| \cdot \cos\theta \qquad (2)$$

where $\theta$ represents the angle between the difference dipole orientation from DFT and the symmetry axis of the CF$_2$ group. Accordingly, we obtain Stark tuning rates from MD-assisted solvatochromism of 0.73 and 0.44 cm$^{-1}$/(MV/cm) for AMOEBA (and similar values for AMBER; see respective entries in brackets in TABLE II), presenting a very good match between the solvatochromic and DFT-based approaches.

The conformational restrictions in DFcH due to the cyclic structure provide an interesting case to discuss the impact of delocalized normal modes on Stark tuning rates and their underlying difference dipole vectors. While the C2-symmetric twisted boat structure is not easily accessible experimentally, our DFT-based analysis in FIG. 3C has shown that this conformation gives rise to conventional $\nu_s$(CF$_2$) and $\nu_{as}$(CF$_2$) modes (again mixed with $\nu$(CC)) with difference dipole moment vectors aligned with the CF$_2$ bisector, i.e. different to the chair conformation. Furthermore, the Stark tuning rate of the former was decreased by a factor of ~3 when comparing to the chair conformation. This exemplifies qualitatively that Stark tuning rates of aliphatic $\nu$(CF) modes are considerably affected by dipolar contributions from the carbohydrate backbone, in the simplest picture the orientation of the C-C dipoles adjacent to the fluorinated unit.

In contrast to DFcH, MFH shows a high degree of flexibility, such that conformation-based changes to the difference dipole become directly relevant. Due to considerable overlap in the experimental spectra of MFH, all individual $\nu$(CF) modes became only accessible via computation. The modes in the range of 1050 – 900 cm$^{-1}$ showed Stark tuning rates of ~0.5 (t1/g1) and ~0.25 cm$^{-1}$/(MV/cm) (t2/g2) with considerably varying difference dipole vectors that were inclined by 9 - 40° from the C-F bond in trans and gauche rotamers of the terminal -CH$_2$F group (FIG. 7B; TABLE II). Like the case of DFcH, this can be understood qualitatively by considering the contribution and orientation of dipoles of the C-C internal coordinates that are part of the normal mode. The only mode that was accessible experimentally without any spectral overlaps was the band at 909 cm$^{-1}$ (g2), which we assigned to gauche conformers of MFH. MD simulations revealed that solvent electrostatics are aligned with the C-F bond dipole: as shown in FIG. 4B, electric field strengths are Gaussian-distributed along the C-F and orthogonal axes, but average fields were negligible only for the latter two axes. Therefore, applying eq. 2 and the angle of the difference dipole from DFT to the solvatochromic slope, we obtain a solvatochromism-based Stark tuning rate of 0.2 cm$^{-1}$/(MV/cm) with AMOEBA fields (0.27 with AMBER), which is overall consistent with the computational results. Despite the considerable spectral overlap, we extracted the spectral peak positions of the modes t1, g1, and t2 by modelling the band pattern between 1030 and 960 cm$^{-1}$ using Gaussian line shapes and report the results in SI section 4 (FIG. S7 and S8). We obtain solvatochromic Stark tuning rate in the range of 0.25 to 0.5 cm$^{-1}$/(MV/cm) with AMOEBA-based fields. The result for t1 differs by a factor of 2 from the DFT-based value, while the other two modes provide a good match (see values in square brackets in TABLE II). However, we want to stress that conformational effects may contribute to these values.

TFH is a similarly conformational flexible case. The two modes in each conformation, i.e. the trans and gauche rotamers of the C-C-C-CF$_3$ dihedral, exhibited considerable electric field-dependent behavior. The t1 and g1 modes, which are based on $\nu_s$(CF$_3$) modes, are characterized by DFT-based tuning rates of 0.21 and 0.33 cm$^{-1}$/(MV/cm), respectively, and considerably tilted vectors by ~80° off the CF$_3$ main axis. Interestingly, an absorption band of the CF$_3$-containing trifluoroethanol at similar frequencies was previously characterized using VSS and revealed a Stark tuning rate of 0.77/$f$ cm$^{-1}$/(MV/cm). Considering $f \approx 2$, this value appears larger than the ones determined herein, which can be attributed to the hydroxyl group which impacts the charge distribution of the molecule and, by this, the Stark tuning rates. A striking result herein is the orientation of the difference dipole vector, which appears to be influenced to a very large degree by the $\delta$(CCC) contribution of the normal mode: the difference dipole points along a CF$_2$ bisector of the CF$_3$ group in the trans rotamer, while in gauche it is rotated by 60° around the C-CF$_3$ bond to point along a C-F bond (FIG. 7 D.1 and D.2). The complicated shape of the absorption feature assigned to this band prevented an experimental solvatochromic analysis.

The t2/g2 modes of TFH that are of $\nu_{as}$(CF$_3$) character showed a clear shift in the solvatochromic spectra. First, DFT suggested that the peak position was hardly affected by transitions between trans and gauche conformations (FIG. 3D and 6B.2), but the difference dipoles varied between 0.49 and 0.29 cm$^{-1}$/(MV/cm) and were inclined by ~60° in different directions from the CF$_3$ main axis. These directions were again along a C-F bond or rotated by 60° toward the adjacent CF$_2$ bisector (FIG. 7D.1 and D.2). However, similar to the above cases of MFH and DFcH, MD solvent electric fields were oriented along the CF$_3$ axis on average. Thus, we applied eq. 2 to obtain solvatochromism-derived Stark tuning rates of 0.82 and/or 0.61 cm$^{-1}$/(MV/cm) according to AMOEBA force fields; note that these values originate from a single experimental solvatochromic slope of 0.34 cm$^{-1}$/(MV/cm) which cannot be separated easily into trans and gauche contributions. Overall, these data exceed the DFT-based value by ~2, which would suggest that AMBER-based results are more accurate, i.e. with values of 0.39 and 0.29 cm$^{-1}$/(MV/cm). However, as discussed in the next section, AMBER (but not AMOEBA) showed considerable deviation in electrostatic parameters, such that we instead propose a different reason for the mismatch. Accordingly, DFT-based spectra are evaluated only in the energy minimum, i.e. at the ideal trans or gauche conformations. However, single bond rotation occurs on the 10 ps time scale due to low barriers of ~4 kcal/mol,[78] such that in both experimental IR spectra and MD simulations conformational fluctuations around the minimum structure are sampled. Since the discussed normal modes and difference dipoles have considerable $\nu$(CF$_3$) and $\delta$(CCC) contributions, dynamics can influence Stark tuning rates in a way that is not grasped by DFT analyses relying on energy minima structures.

This effect may be observable for TFH, because the 1150 cm$^{-1}$ band originates from an overlap of trans and gauche rotamers, but not for MFH, where the 909 cm$^{-1}$ band was solely due to gauche conformations.

## B. Differences between results from AMOEBA and AMBER force fields

The MD-assisted solvatochromic analysis revealed consistent results for MFH and DFcH using AMOEBA and AMBER-based electric fields, but for FB and TFH discrepancies by factors of 0.5 to 2 have been observed (TABLE II), which can be directly related to differences in the extracted electric field strengths (FIG. 4A, 4D). A reason for these discrepancies between results from AMBER and AMOEBA can be found in the electrostatic parameters in the force fields. Whereas the generalized AMBER force field (GAFF) only assigns a partial charge to each atom, each atom in AMOEBA´s force field carries charge, dipole, quadrupole, and polarizability parameters to enable a more accurate description of the electrostatic potential. This is evident when evaluating the deviation of electrostatic potentials based on the force fields and quantum mechanical (QM) calculations (MP2/aug-cc-pvtz). For all four solutes (see full TABLE S3 and S4), the RMSD between the electrostatic potential from AMOEBA and QM was 0.08 kcal/mol/e, which corresponded to relative deviations by 1 – 3%. In turn, the AMBER-based electrostatic potential deviated by ten times more from QM, i.e., by RMSDs 0.7 - 1.0 kcal/mol/e or 16 - 35%. Overall, this points to a more reliable description of the electrostatics of fluorinated compounds using the AMOEBA force field.

When specifically analyzing the RMSD on the atoms contributing to the $CF_n$ groups, we find that the C and F atom(s) in MFH and DFcH show RMSDs between 0.5 and 1.1 kcal/mol/e, which is within the average range of the deviations. Instead, RMSDs per C and F atom(s) are 2.17 and 0.74 kcal/mol/e in FB, and the electrostatic potential deviates by 2.8 and 1.2 kcal/mol/e in TFH. Thus, we infer that all F atoms are reasonably described within the average RMSD of the AMBER's parameters of the entire molecules. However, the electrostatics due to the C atoms is much more error-prone, affecting the local electrostatic potential and, by this, the organization of the solvent around the $CF_n$ groups. Overall, this suggests that force fields containing higher-order multipoles, such as AMOEBA, can be necessary to accurately describe the electrostatic interaction of fluorinated solutes with solvents or fluorous interactions overall.

## V. CONCLUSIONS

Previous work focused mainly on the aromatic v(CF) of FB as a VSE probe.[42–44] The collected results on aromatic and aliphatic fluorinated compounds provided herein enable us to draw general conclusions on the use of $CF_n$ groups for VSE analyses.

- Overall, we can conclude that the Stark tuning rates of the vibrational modes that were experimentally accessible, vary in the range of 0.2 – 0.8 cm$^{-1}$/(MV/cm), similar to those found for v(C≡N) and v(C=O) probes.[21–24,45,46] Within this range, the smallest values were observed for $CF_{aliphatic}$, whereas the largest electric-field sensitivity was found for $CF_{3,aliphatic}$ (a general trend of $CF_{aliphatic} \leq CF_{aromatic} < CF_{2,aliphatic} \leq CF_{3,aliphatic}$ is found for the averaged values; see FIG. S9). This is generally in line with the increasing dipole moment of the $CF_n$ unit with polyfluorination.

- While the common VSE probes, like v(C≡N) and v(C=O), report on the electric field along the bond axis, this is only the case for the aromatic v(CF) mode. In the case of the aliphatic probes, the difference dipole vectors were tilted away from the main symmetry axes of the $CF_n$ unit by up to 70° for the experimentally accessible probes. The orientation of the vector is further dependent on the molecular conformation adapted and, therefore, must be considered when using aliphatic v(CF) probes for VSE analyses. The aromatic v(CF) mode presents the exception to this effect due to its conformational constraints of the phenyl unit.

- While the tilted direction of the difference dipole with respect to the $CF_n$ symmetry axes adds a level of complexity to the analysis of electric fields in condensed media, it can provide the opportunity to quantify electrostatics in several dimensions. Similar to the previously demonstrated deuterated aldehydes (i.e. v(CO) and v(CD) probes in one functional group),[49] the $CF_2$ (and to some degree the $CF_3$) groups provide VSE sensitive vibrational modes that provide access to electric fields along the distinct directions of the difference dipoles for given conformations.

- Finally, we note that solvatochromic calibrations of novel molecules with $CF_n$ probes as performed herein can require the usage of computational approaches of a higher level of theory as conventional fixed-charge MD force fields can encounter issues with accurate modeling of the electrostatic properties of fluorinated compounds. Classical simulations that take advantage of including higher order multipoles and polarizability, as in the AMOEBA force fields, model the electrostatics of $CF_n$ groups to higher accuracy.

With the relevance of fluorination as a strategy to alter molecular properties, establishing physical experimental methods to quantify the effects on non-covalent interactions due to derivatization is highly desirable. Interpreting vibrational spectra using the framework of the VSE is a particularly suitable approach towards this goal as exemplified by the numerous existing studies of C≡N and C=O probes in solution,[21–26] at metal interfaces[27–29] or within protein or lipid environments[30–39]. The present work provides a guide to using the simplest aliphatic (and aromatic) $CF_n$ motifs in such studies to rationally fine-tune fluorous interactions for many medicinal and biophysical applications. This work further aims at inspiring future studies of more complex fluorination patterns that may require non-linear spectroscopy and/or high-level theory to dissect the relation between vibrational spectra and the underlying structural and electrostatic properties.

## ACKNOWLEDGMENTS

The authors thank the HPC Service of ZEDAT, Freie Universität Berlin, for computing time (10.17169/refubium-26754) and the German Research Foundation (DFG) for financial support via the Sonderforschungsbereich 1349 (SFB 1349) "Fluor-Specific Interactions: Fundamentals and Functions" (project number 387284271 – project C05, J.H.) and the DFG Individual Research Grant KO 5464-4 (project number 493270578, J.K.). Molecular structures were created using the matlab scripts molecule3D.m[79] (André Ludwig) and Arrow3.m[80] (Tom Davis).

## CORRESPONDING AUTHORS

**Joachim Heberle:** joachim.heberle@fu-berlin.de
orcid.org/0000-0001-6321-2615

**Jacek Kozuch:** jacek.kozuch@fu-berlin.de
orcid.org/0000-0002-2115-4899

## AUTHOR CONTRIBUTION

**Ruben Cruz:** Experiments (main); Simulations (main); Formal analysis (main); Methodology (equal); Writing – original draft (equal); Writing – review & editing (equal); **Kenichi Ataka:** Supervision (equal); Writing – review & editing (equal); **Joachim Heberle:** Conceptualization (equal); Supervision (equal); Resources (equal); Writing – review & editing (equal); **Jacek Kozuch:** Conceptualization (equal); Simulations (supporting); Methodology (equal); Supervision (equal); Resources (equal); Writing – original draft (equal); Writing – review & editing (equal).

# Supporting Information for

# Evaluating aliphatic CF, CF$_2$ and CF$_3$ groups as vibrational Stark effect reporters

R. Cruz,[1] K. Ataka,[1] J. Heberle,[1,2,*] and J. Kozuch [1,2,*]

[1] *Fachbereich Physik, Freie Universität Berlin, Berlin, 14195,* Germany

[2] *Forschungsbau SupraFAB, Freie Universität Berlin, Berlin, 14195,* Germany

**\* Joachim Heberle:** joachim.heberle@fu-berlin.de
**\* Jacek Kozuch:** jacek.kozuch@fu-berlin.de

# Contents

1. Electric Field Distributions in MD Simulations

In order to determine the orientation of solvation electric fields on the $CF_n$ groups in MD simulations, we first introduced a local reference frame as specified in FIG. S1. Electric fields were determined as the average electric field on the C and F atom(s) of the $CF_n$ groups and projected onto the axes of the reference frame. FIG. S2-S6 depict the histograms of the electric field along the MD trajectories, showing that average field projections along x and y are negligible, and the electric fields along the z-axes dominate. These electric field orientations match the overall static dipole of the $CF_n$ groups.

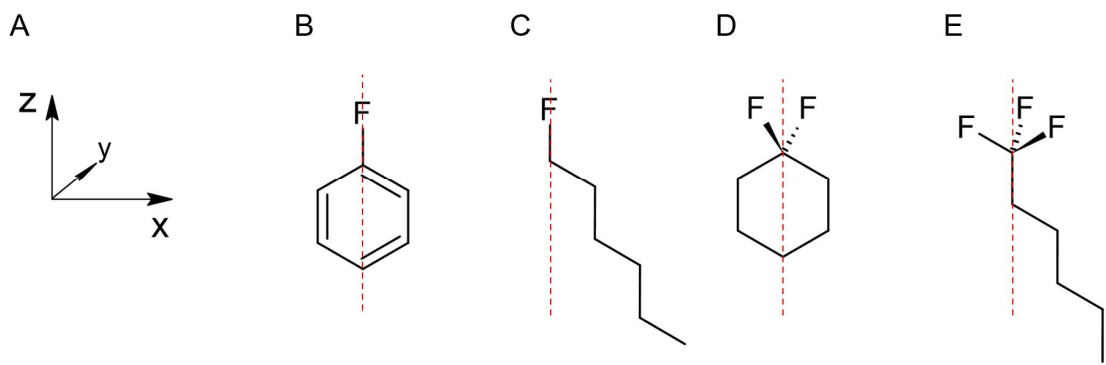

FIG. S1: Local frame definition for the determination of the directions of electric fields in MD simulations as specified by the coordinate axes (A) and the molecules FB, MFH, DFcH, and TFH (B-E). The dashed red line indicates the z-axis; the y-axis is pointing into the plane. The z-axis of aromatic and aliphatic CF groups (B and C, respectively) is the C-F bond axis and the x-axis is given by the C2 atom. The z-axis of the $CF_2$ group (D) is the FCF bisector and the x-axis is defined by the C2 atom. The z-axis of the $CF_3$ group (E) is the $C_3$ symmetry axis of the and the x-axis is defined by the C3 atom.

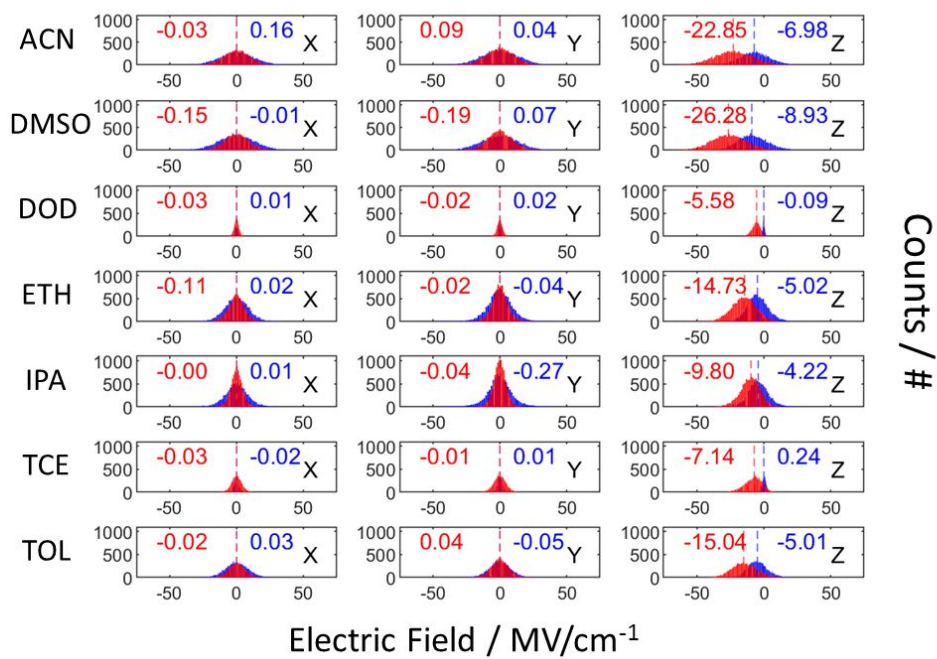

FIG. S2: Electric field distributions on the CF group of fluorobenzene (FB) along x, y, and z as defined in FIG. S1. Solvents from top to bottom are arrange in alphabetical order (ACN = acetonitrile, DMSO = dimethylsulfoxide, DOD = n-dodecane, ETH = ethanol, IPA = isopropanol, TCE = chloroform, TOL = toluene). Red and blue distributions originate from AMOEBA and AMBER, respectively.

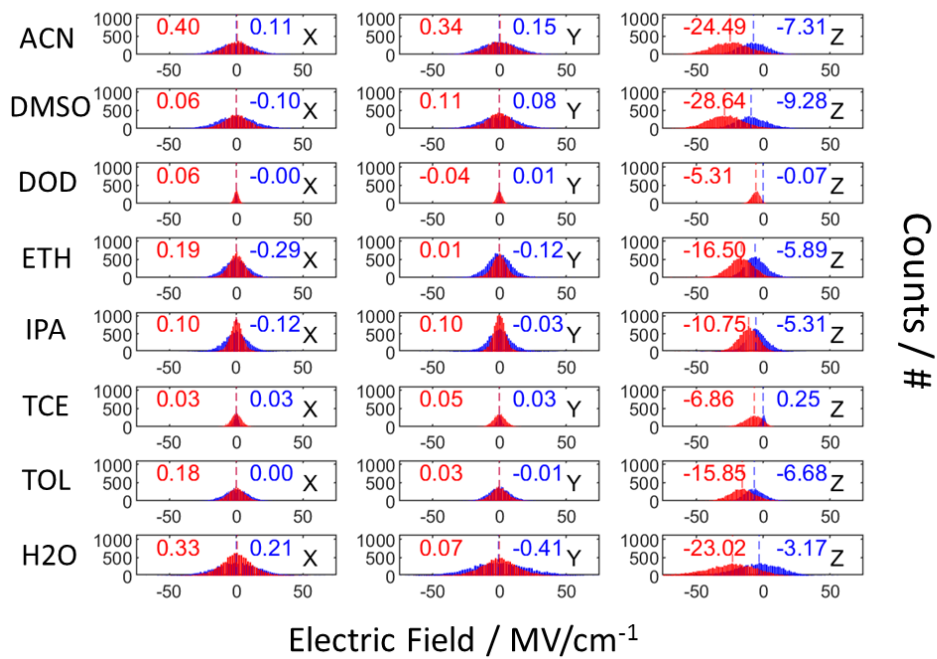

FIG. S3: Electric field distributions on the CF group of 4-fluorobenzyl alcohol (FBA) along x, y, and z as defined in FIG. S1. Solvents from top to bottom are arrange in alphabetical order (ACN = acetonitrile, DMSO = dimethylsulfoxide, DOD = n-dodecane, ETH = ethanol, IPA = isopropanol, TCE = chloroform, TOL = toluene). Red and blue distributions originate from AMOEBA and AMBER, respectively.

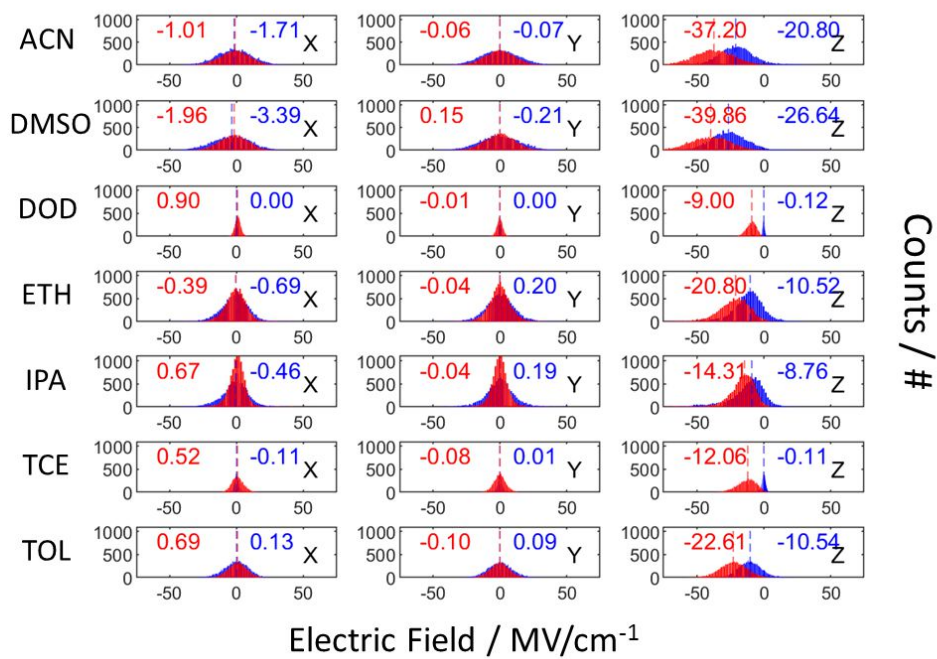

FIG. S4: Electric field distributions on the CF group of 1-(mono)fluorohexane (MFH) along x, y, and z as defined in FIG. S1. Solvents from top to bottom are arrange in alphabetical order (ACN = acetonitrile, DMSO = dimethylsulfoxide, DOD = n-dodecane, ETH = ethanol, IPA = isopropanol, TCE = chloroform, TOL = toluene). Red and blue distributions originate from AMOEBA and AMBER, respectively.

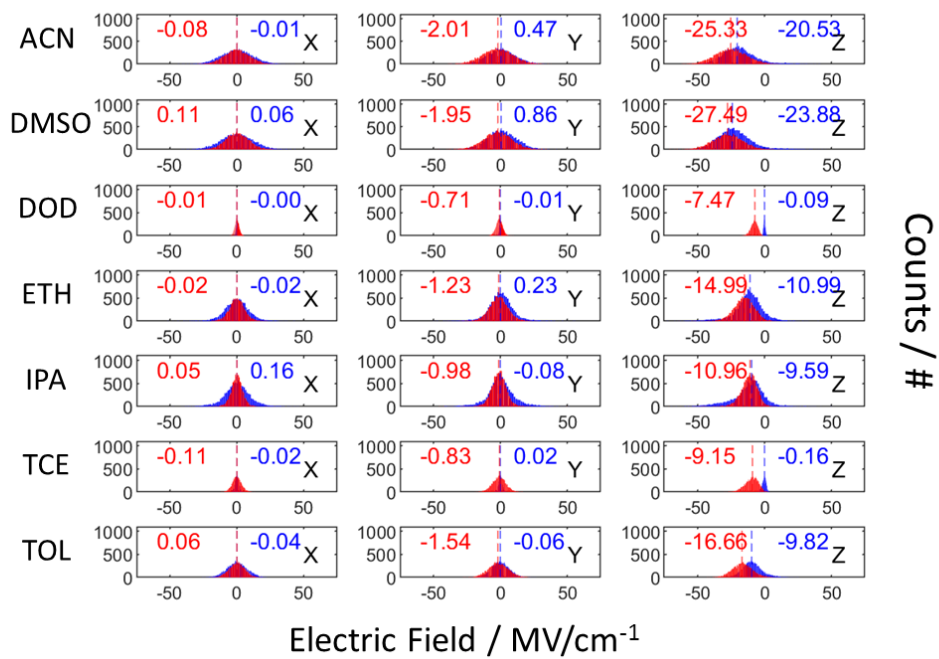

FIG. S5: Electric field distributions on the $CF_2$ group of 1,1-difluorocylcohexane (DFcH) along x, y, and z as defined in FIG. S1. Solvents from top to bottom are arrange in alphabetical order (ACN = acetonitrile, DMSO = dimethylsulfoxide, DOD = n-dodecane, ETH = ethanol, IPA = isopropanol, TCE = chloroform, TOL = toluene). Red and blue distributions originate from AMOEBA and AMBER, respectively.

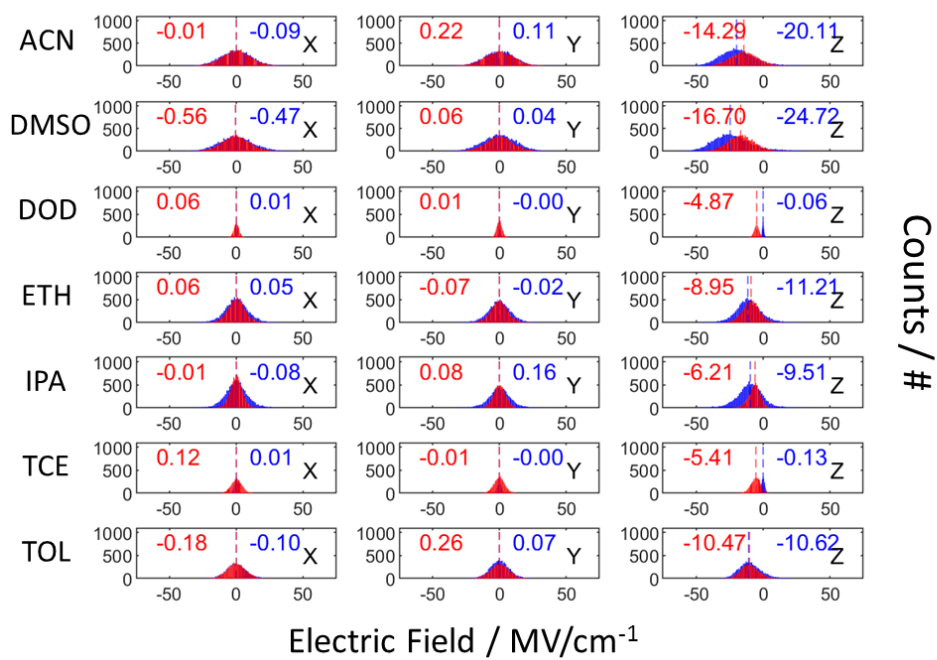

FIG. S6: Electric field distributions on the $CF_2$ group of 1,1,1-trifluorohexane (TFH) along x, y, and z as defined in FIG. S1. Solvents from top to bottom are arrange in alphabetical order (ACN = acetonitrile, DMSO = dimethylsulfoxide, DOD = n-dodecane, ETH = ethanol, IPA = isopropanol, TCE = chloroform, TOL = toluene). Red and blue distributions originate from AMOEBA and AMBER, respectively.

## 2. DFT-based Stark Parameters

TABLE S1: DFT-based Stark parameters resulting from fitting equation 1 (main text) to the vibrational wavenumbers obtained after normal mode analysis in external electric fields (FIG. 3; see materials and methods section in main text for details).

| Molecule | Wavenumber cm$^{-1}$ | $\mu_x$ | $\mu_y$ cm$^{-1}$/(MV/cm) | $\mu_z$ | $\alpha_{xx}$ | $\alpha_{yy}$ cm$^{-1}$/(MV/cm)$^2$ | $\alpha_{zz}$ | $R^2$ |
|---|---|---|---|---|---|---|---|---|
| FB | 1234.7 | 0.00 | 0.00 | -0.63 | 0.0003 | 0.0002 | 0.0001 | 0.999 |
| FBA | 1239.4 | 0.01 | 0.07 | -0.74 | 0.0002 | 0.0011 | 0.0006 | 0.999 |
| MFH gauche | 1001.8 | 0.00 | 0.10 | -0.20 | 0.0015 | 0.0013 | 0.0023 | 0.992 |
|  | 922.9 | 0.06 | -0.15 | -0.20 | 0.0005 | 0.0011 | 0.0017 | 0.996 |
| MFH trans | 1023.3 | 0.07 | 0.00 | -0.48 | 0.0020 | 0.0014 | -0.0018 | 0.996 |
|  | 988.0 | 0.15 | 0.00 | -0.49 | 0.0027 | 0.0044 | 0.0080 | 0.997 |
| DFcH chair | 956.3 | 0.00 | 0.12 | -0.40 | 0.0003 | 0.0054 | 0.0013 | 0.999 |
|  | 1110.2 | 0.00 | -0.61 | -0.44 | 0.0006 | -0.0005 | 0.0006 | 0.999 |
| DFcH twisted boat | 945.1 | 0.00 | 0.00 | -0.40 | 0.0002 | 0.0102 | 0.0002 | 0.993 |
|  | 1100.4 | 0.00 | 0.00 | -0.26 | 0.0005 | -0.0043 | 0.0008 | 0.998 |
| TFH gauche | 1149.4 | 0.23 | 0.08 | -0.16 | 0.0015 | -0.0005 | -0.0002 | 0.988 |
|  | 1253.5 | -0.33 | -0.02 | -0.03 | -0.0020 | 0.0016 | 0.0009 | 0.997 |
| TFH trans | 1148.6 | 0.45 | 0.00 | -0.20 | -0.0050 | 0.0014 | -0.0003 | 1.000 |
|  | 1256.3 | -0.21 | 0.00 | -0.04 | 0.0011 | -0.0053 | 0.0014 | 0.998 |

## 3. MD-associated Solvatochromism Parameters

TABLE S2: Fitting parameters to the vibrational wavenumbers obtained from MD-assisted solvatochromism-based analysis (FIG. 5, 6; see materials and methods section in main text for details). Results in brackets result from overlapping bands that are evaluated in SI section 4.

|  | Wavenumber (TCE) cm$^{-1}$ | Slope cm$^{-1}$/(MV/cm) | Intercept cm$^{-1}$ | $R^2$ | Slope cm$^{-1}$/(MV/cm) | Intercept cm$^{-1}$ | $R^2$ |
|---|---|---|---|---|---|---|---|
|  |  | *AMBER* |  |  | *AMOEBA* |  |  |
| FB | 1222 | 0.77 | 1221.7 | 0.60 | 0.38 | 1224.5 | 0.96 |
| MFH | (1015) | (0.30 | 1017.9 | 0.82) | (0.27 | 1020.6 | 0.98) |
|  | (1006) | (0.56 | 1006.7 | 0.82) | (0.50 | 1011.4 | 0.95) |
|  | (996) | (0.31 | 982.6 | 0.90) | (0.27 | 985.0 | 0.99) |
|  | 916 | 0.21 | 916.2 | 0.90 | 0.15 | 917.1 | 0.75 |
| DFcH | 1113 | 0.40 | 1114.2 | 0.86 | 0.43 | 1117.1 | 0.84 |
|  | 961 | 0.34 | 962.6 | 0.63 | 0.42 | 965.8 | 0.98 |
| TFH | 1150 | 0.16 | 1150.5 | 0.62 | 0.34 | 1151.9 | 0.68 |

## 4. Evaluation of Overlapping v(C-F) Bands of Monofluorohexane

To provide an estimate of MFH's solvatochromic Stark tuning rate of the overlapping IR absorption bands at 1020, 990, and 980 cm$^{-1}$ (referred to as t1, g1, and t2, respectively), we modelled the complicated band shape with three Gaussian components as shown in FIG. S7 A (the spectrum of MFH in DODEC required a fourth broad band, which may originate from background). Since MFH in TOL and EtOH shows a strong negative band at 1050 cm$^{-1}$ originating from solvent absorption, we further provide the second derivative spectra in FIG. S7 B in order to provide confidence to the peak positions. We do not show the spectrum in DMSO in this spectral region, due to strong overlap with solvent bands. Overall, we observe a consistent redshift by 9, 14, and 8 cm$^{-1}$ for the bands that we assign to the t1, g1, and t2, respectively. Plotting the peak positions with MD-based electric fields along the C-F bond of MFH (FIG. S8 A, B, C), we obtain linear correlations with $R^2$ of 0.97 – 0.99 for AMOEBA and 0.82 – 0.9 for AMBER. The resulting solvatochromic slopes are 0.27, 0.50, and 0.27 cm$^{-1}$/(MV/cm) for t1, g1, and t2 for AMOEBA; the corresponding slopes with AMBER force fields are 0.3, 0.56, and 0.31 cm$^{-1}$/(MV/cm). Despite the consistent linear trends and (very) good quality fits, these slopes may contain contributions from the conformational drift across solvents (see discussion in main text).

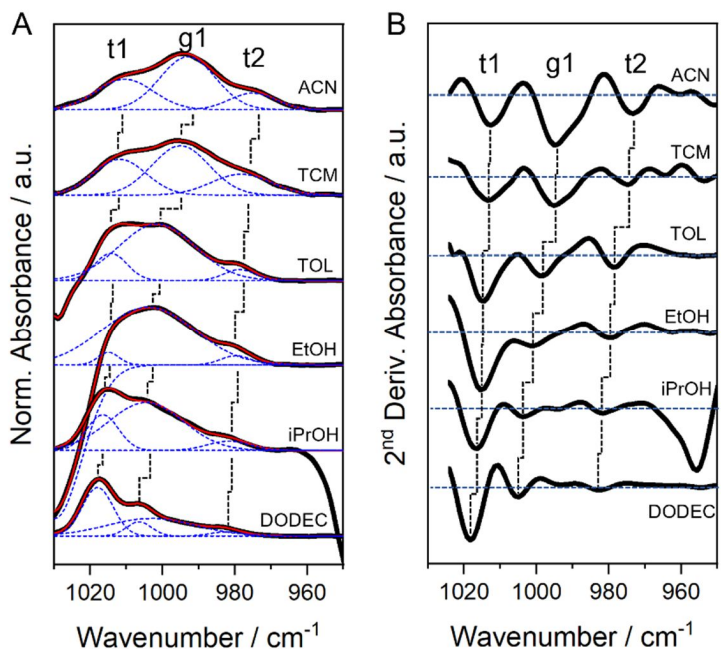

FIG. S7: Solvatochromic spectra (A) and second derivative spectra (B) of MFH (solvents specified in the figure) in the spectral region of 1030 – 950 cm$^{-1}$; spectra and second derivatives are stacked and normalized. A: Solvatochromic spectra were modelled with three Gaussian band shapes to extract the peak positions of the bands t1, g1, and t2 (see assignment in the main text). In TOL and EtOH a negative band at < 1020 cm-1 originates from the solvent and was modelled with an additional Gaussian of negative intensity. In DODEC a fourth Gaussian components yields a broad band, which can be assigned to a background signal. B: Second derivative spectra were calculated from spectra in A, reproducing the peak positions from Gaussian band shape fits. Horizontal dashed lines show the zero signal line in each second derivative spectrum.

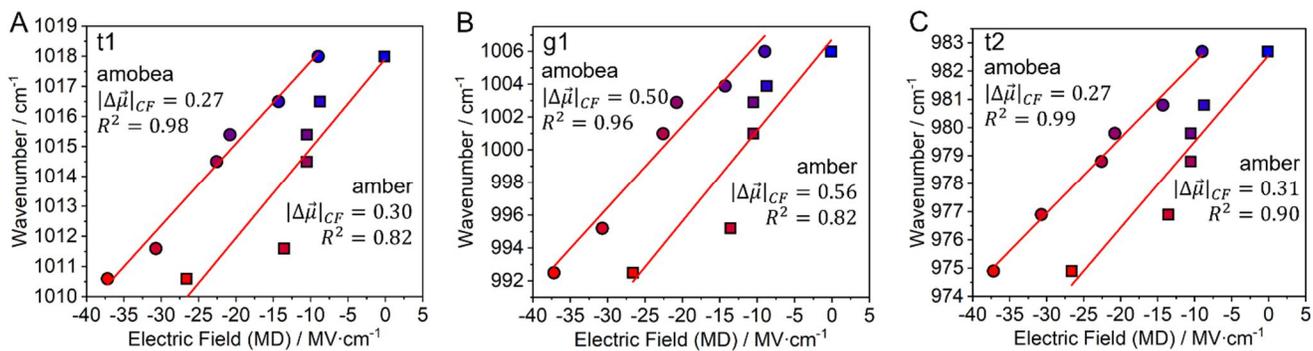

FIG. S8: Correlation of peak positions and MD-based electric fields for t1 (A), g1 (B), and t2 (C) modes of MFH. Data point color specifies the solvent used, i.e. dodecane, i-propanol, ethanol, toluene, chloroform, and acetonitrile along the gradient from blue to red.

# 5. Electrostatic Potential Comparison of AMBER, AMEOBA and QM

TABLE S3: Comparison of atomic electrostatic potentials (all entries are in units of kcal/e/mole) of fluorobenzene (FB) from a quantum mechanical (QM) calculation (MP2/6-311++g**) with parameters from AMBER and AMOEBA. There relevant C and F atoms are highlighted in bold. Overall difference of AMBER-QM and AMOEBA-QM are 0.9685 and 0.0791, respectively.

| Atom | ESP QM | ESP AMBER | RSMD AMBER-QM | ESP AMOEBA | RMSD AMOEBA-QM |
|---|---|---|---|---|---|
| **C** | **-4.8157** | **-6.8645** | **2.1743** | **-4.8851** | **0.1375** |
| **F** | **-5.7757** | **-6.332** | **0.7407** | **-5.7997** | **0.0723** |
| C | -2.4346 | -3.2186 | 1.0448 | -2.4614 | 0.098 |
| C | -1.4849 | -0.8637 | 0.911 | -1.5174 | 0.0853 |
| C | -1.4444 | -0.3638 | 1.2893 | -1.4819 | 0.0973 |
| C | -1.5024 | -0.8978 | 0.9254 | -1.5402 | 0.0903 |
| C | -2.4378 | -3.2249 | 1.0462 | -2.4685 | 0.1002 |
| H | 1.7959 | 0.976 | 0.8959 | 1.7592 | 0.0633 |
| H | 4.013 | 4.5288 | 0.6503 | 3.9804 | 0.0552 |
| H | 3.9553 | 4.9365 | 1.021 | 3.918 | 0.0744 |
| H | 4.0089 | 4.5207 | 0.6479 | 3.9729 | 0.058 |
| H | 1.7944 | 0.9748 | 0.8951 | 1.7545 | 0.0651 |

TABLE S4: Comparison of atomic electrostatic potentials (all entries are in units of kcal/e/mole) of 1-(mono)fluorohexane (MFH; 6 conformations used in total) from a quantum mechanical (QM) calculation (MP2/6-311++g**) with parameters from AMBER and AMOEBA. There relevant C and F atoms are highlighted in bold. Overall difference of AMBER-QM and AMOEBA-QM are 0.9637 and 0.0791, respectively.

| Atom | ESP QM | ESP AMBER | RSMD AMBER-QM | ESP AMOEBA | RMSD AMOEBA-QM |
|---|---|---|---|---|---|
| **C** | **1.2938** | **2.207917** | **1.107133** | **1.214583** | **0.112233** |
| **F** | **-8.23483** | **-7.75942** | **0.824933** | **-8.25923** | **0.0847** |
| H | 2.50595 | 2.513033 | 0.617117 | 2.474317 | 0.070733 |
| H | 2.48695 | 2.5433 | 0.616617 | 2.455483 | 0.071667 |
| C | 2.423017 | 3.49575 | 1.232783 | 2.3483 | 0.1497 |
| H | 1.28765 | 0.769183 | 1.048883 | 1.241667 | 0.1078 |
| H | 3.355633 | 3.298017 | 0.8098 | 3.33965 | 0.078017 |
| C | -1.33398 | -0.95965 | 1.053233 | -1.36352 | 0.10805 |
| H | 0.76225 | -0.1252 | 1.200983 | 0.771167 | 0.076083 |
| H | -0.24208 | -1.22998 | 1.331267 | -0.24648 | 0.075 |
| C | 0.9773 | 2.046933 | 1.677317 | 0.908683 | 0.118267 |
| H | 2.337617 | 1.713033 | 1.209667 | 2.315933 | 0.066717 |
| H | 2.031 | 1.278317 | 1.3194 | 2.00635 | 0.068333 |
| C | -0.00177 | 0.262933 | 0.737417 | -0.045 | 0.097567 |
| H | 1.396117 | 0.775617 | 0.88515 | 1.378317 | 0.061817 |
| H | 0.563733 | -0.29013 | 1.15165 | 0.531417 | 0.068967 |
| C | -0.01692 | 0.9331 | 1.2329 | -0.06648 | 0.09025 |
| H | 0.872533 | 0.8611 | 0.68585 | 0.838383 | 0.064683 |
| H | 0.89285 | 0.848933 | 0.7445 | 0.861933 | 0.061383 |
| H | 1.1501 | 1.18585 | 0.753317 | 1.12505 | 0.057117 |

TABLE S5: Comparison of atomic electrostatic potentials (all entries are in units of kcal/e/mole) of 1,1-difluorocyclohexane (DFcH) from a quantum mechanical (QM) calculation (MP2/6-311++g**) with parameters from AMBER and AMOEBA. There relevant C and F atoms are highlighted in bold. Overall difference of AMBER-QM and AMOEBA-QM are 0.6968 and 0.0652, respectively.

| Atom | ESP QM | ESP AMBER | RSMD AMBER-QM | ESP AMOEBA | RMSD AMOEBA-QM |
|---|---|---|---|---|---|
| **C** | **0** | **0** | **0** | **0** | **0** |
| **F** | **-7.8566** | **-7.6773** | **0.5386** | **-7.864** | **0.0619** |
| **F** | **-7.9135** | **-8.8428** | **0.985** | **-7.9674** | **0.0829** |
| C | 1.9362 | 2.6822 | 0.8403 | 1.8886 | 0.1053 |
| H | -0.3518 | -0.8242 | 0.773 | -0.3869 | 0.0702 |
| H | 3.4404 | 3.5932 | 0.4467 | 3.405 | 0.072 |
| C | 2.6239 | 3.2274 | 0.7569 | 2.5643 | 0.0978 |
| H | 0.7251 | 0.1802 | 0.878 | 0.7056 | 0.0486 |
| H | 4.0626 | 4.388 | 0.4382 | 4.0518 | 0.0501 |
| C | 4.2602 | 5.5177 | 1.3438 | 4.2161 | 0.0701 |
| H | 5.2241 | 5.6052 | 0.6436 | 5.2153 | 0.0368 |
| H | 3.861 | 4.1685 | 0.5986 | 3.8355 | 0.0599 |
| C | 2.7101 | 3.3255 | 0.7692 | 2.6502 | 0.0987 |
| H | 4.0476 | 4.3738 | 0.4393 | 4.0376 | 0.0491 |
| H | 0.7395 | 0.1885 | 0.8695 | 0.7207 | 0.0471 |
| C | 2.0051 | 2.7398 | 0.8324 | 1.9591 | 0.1033 |
| H | 3.4268 | 3.5839 | 0.451 | 3.3916 | 0.0727 |
| H | -0.3086 | -0.7801 | 0.7717 | -0.3443 | 0.0713 |

TABLE S6: Comparison of atomic electrostatic potentials (all entries are in units of kcal/e/mole) of 1,1,1-trifluorohexane (TFH; 6 conformations used in total) from a quantum mechanical (QM) calculation (MP2/6-311++g**) with parameters from AMBER and AMOEBA. There relevant C and F atoms are highlighted in bold. Overall difference of AMBER-QM and AMOEBA-QM are 1.0948 and 0.0638, respectively.

| Atom | ESP QM | ESP AMBER | RSMD AMBER-QM | ESP AMOEBA | RMSD AMOEBA-QM |
|---|---|---|---|---|---|
| **C** | **-7.81262** | **-10.3106** | **2.773867** | **-7.79878** | **0.070583** |
| **F** | **-5.00868** | **-5.92683** | **1.202617** | **-5.03638** | **0.06345** |
| **F** | **-5.00218** | **-5.71665** | **1.138833** | **-5.02657** | **0.06265** |
| **F** | **-5.02402** | **-6.14495** | **1.265983** | **-5.05093** | **0.062017** |
| C | 4.762267 | 7.659383 | 2.9963 | 4.6889 | 0.103 |
| H | 4.313167 | 5.1017 | 1.175333 | 4.2912 | 0.061 |
| H | 4.324667 | 5.11415 | 1.17485 | 4.3049 | 0.059833 |
| C | 0.403583 | 1.613467 | 1.42955 | 0.36315 | 0.089917 |
| H | 1.771233 | 1.569033 | 0.646933 | 1.77145 | 0.0558 |
| H | 1.773933 | 1.56725 | 0.641017 | 1.77425 | 0.0555 |
| C | 4.8499 | 7.4148 | 2.8014 | 4.8105 | 0.097367 |
| H | 5.372833 | 5.866267 | 1.121317 | 5.359317 | 0.060867 |
| H | 5.3698 | 5.86875 | 1.11995 | 5.355917 | 0.061383 |
| C | 2.283467 | 3.3982 | 1.34135 | 2.254233 | 0.086917 |
| H | 3.09035 | 3.18655 | 0.593333 | 3.070567 | 0.063467 |
| H | 3.0927 | 3.190317 | 0.598683 | 3.072883 | 0.0641 |
| C | 2.0085 | 3.507117 | 1.664567 | 1.96225 | 0.09095 |
| H | 3.11275 | 3.719083 | 0.9104 | 3.092867 | 0.05545 |
| H | 2.982067 | 3.567033 | 0.869567 | 2.957783 | 0.05945 |
| H | 2.8585 | 3.418867 | 0.823683 | 2.8305 | 0.062567 |

## 6. Trend of Solvatochromic Tuning Rates

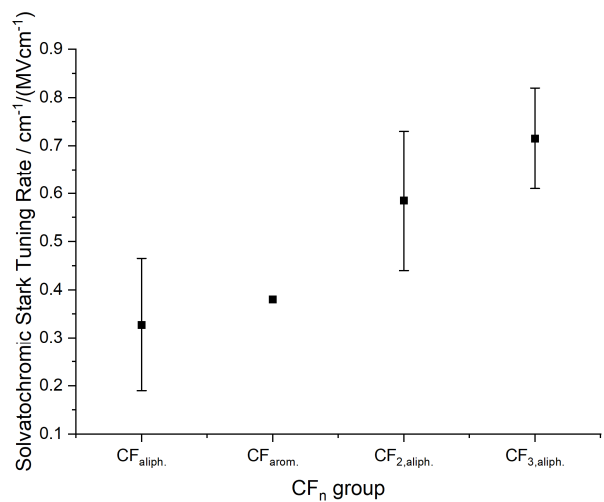

FIG. S9: Trend of average solvatochromic Stark tuning rates from TABLE II from the main text with increasing number of F atoms in the $CF_n$ group. Bars are standard deviations of the values obtained from different vibrational modes.